# Applications of Belief Propagation in CSMA Wireless Networks

Cai Hong Kai, Soung Chang Liew
Department of Information Engineering, The Chinese University of Hong Kong
Email: {chkai6, soung}@ie.cuhk.edu.hk

*Abstract*— **The belief propagation (BP) algorithm is an efficient way to solve "inference" problems in graphical models, such as Bayesian networks and Markov random fields. The system-state probability distribution of CSMA wireless networks is a Markov random field. An interesting question is how BP can help the analysis and design of CSMA wireless networks. This paper explores three such applications. First, we show how BP can be used to compute the throughputs of different links in the network given their access intensities, defined as the mean packet transmission time divided by the mean backoff countdown time. Second, we propose an inverse-BP algorithm to solve the reverse problem: how to set the access intensities of different links to meet their target throughputs? Third, we introduce a BP-adaptive CSMA algorithm to find the link access intensities that can achieve optimal system utility. BP solves the three problems with exact results in networks with tree contention graph. It may, however, lose accuracy in networks with a loopy contention graph. We then show how a generalized version of BP, GBP, can be designed to solve the three problems with high accuracy for networks with loopy contention graph. Importantly, we show how the BP and GBP algorithms in this paper can be implemented in a distributed manner, making them useful in practical CSMA network operation.**

## I. INTRODUCTION

With the widespread deployment of IEEE 802.11 networks, it is common today to find multiple wireless LANs co-located in the neighborhood of each other. The multiple wireless LANs form an overall large network whose links interact and compete for airtime using the carrier-sense multiple access (CSMA) protocol. The carrier sensing relationships among these links are often "non-all-inclusive" in that each link may sense only a subset, but not all, of other links.

For analytical purposes, the carrier sensing relationships among the links are typically captured using a contention graph. The links are modeled by vertices of the graph, and an edge joins two vertices if the transmitters of the two associated links can sense each other. Since different links may sense different subsets of other links, the links may experience different throughputs in the network.

Ref. [1] presented an analytical model, Ideal CSMA Network (ICN), to study the behavior of CSMA networks given their contention graphs. It was shown that the throughputs of links can be computed from the stationary probability distribution of the states of a continuous-time Markov chain. Furthermore, the contention graph associated with ICN is a Markov random field [2] with respect to the probability distribution of its system states. The belief propagation (BP) algorithm is an efficient way to solve "inference" problems in graphical models, such as Bayesian networks and Markov random fields [3]. An interesting question, therefore, is how BP can help the analysis and design of CSMA wireless networks.

This paper considers three applications of BP in CSMA networks. To the best of our knowledge, this is the first paper to use the BP framework to solve problems related to CSMA networks. Importantly, we show that all three problems are amenable to solutions by distributed algorithms under the BP framework.

The first and the most direct application is to use BP to compute (infer) the throughputs of different links in a CSMA network given their access intensities. The access intensity of a link is the ratio of its average packet transmission time to its average backoff countdown time. Higher access intensity corresponds to higher aggressiveness of the link when it competes for airtime under the CSMA protocol. BP gives exact solutions for tree contention graphs and acceptable approximate solutions for loopy contention graphs. We show that an improved algorithm, generalized belief algorithm (GBP), can reduce the errors induced by loops significantly.

The second application is the reverse problem of computing the link access intensities to meet the target link throughputs. We propose an Inverse Belief Propagation (IBP) algorithm for this purpose. IBP can quickly output the approximate link access intensities required. Analogous to GBP, we propose IGBP to reduce the errors in the access intensities found.

The third application is on network utility optimization. We propose a BP-adaptive CSMA algorithm (BP-ACSMA) to adaptively achieve the optimal system utility. Compared with prior work, an advantage of BP-ACSMA is that it is a *proactive* computational algorithm without the need for network probing and traffic measurement. As with GBP and IGBP, we propose GBP-ACSMA for higher accuracy in loopy graphs. Our simulation results indicate that the achieved aggregate throughputs and system utility are near optimal.

### Related Work

There have been numerous publications on non-all-inclusive carrier-sense networks and this is indeed a "hot topic" among researchers. Recent work includes [1], [4]-[6], from which earlier work can be traced. Among them, [1] proposed a quick "back-of-the-envelope" (BoE) algorithm for link throughputs computation in CSMA wireless networks. BoE could handle networks of up to 50 links with high accuracy and speed. Networks of larger size were left as an open issue. The BP algorithm proposed in this paper fills this gap.

Besides throughput computation, this paper proposes and investigates two other applications of BP: (1) computation of link access intensities required to meet target link throughputs; (2) optimization of network utility. The existing algorithms proposed in [7] are based on "probe and measure". Specifically, before a link adjusts its access intensity, a period of "smoothing" time is needed to measure the difference in the link's input traffic and output traffic. As will be shown in this paper, the required smoothing time can be quite excessive in networks that exhibit temporal starvation, resulting in very slow convergence. By contrast, BP-based algorithms proposed here do not have this problem because they are computation-based rather than measurement-based.

BP as an inference-making methodology has been studied extensively. A good reference for BP is [8]. We believe ours is the first paper to explore the applications of BP in CSMA networks. Ref. [8] also presents GBP, without focusing on specific application domains. An important contribution of our paper is to show that a "maximal clique" method of forming "regions" in GBP allows us to design adaptive *distributed* GBP algorithms for CSMA networks. Furthermore, this region-forming method yields good performance.

**Paper Organization**

The remainder of the paper is organized as follows. Section II introduces our system model and reviews the throughput computation of CSMA wireless networks. Section III shows how to use BP for throughput computation in large CSMA wireless networks. Section IV investigates the reverse problem: given the target link throughputs, how to find the link access intensities to meet them. Section V proposes the BP-ACSMA algorithm for network utility optimization. Section VI shows how GBP can be used to solve the same problems as in Sections III-V, but with higher accuracy. Section VII concludes this paper.

## II. SYSTEM MODEL

In this section, we first review an idealized version of the CSMA network (ICN) to capture the main features of the CSMA protocol responsible for the interaction and dependency among links. The ICN model was used in several prior investigations [1][4][5][7]. The correspondence between ICN and the IEEE 802.11 protocol [9] can be found in [1].

### A. The ICN model

In ICN, the carrier-sensing relationship among links is described by a contention graph $G = (V, E)$ [^1]. Each link is modeled as a vertex $i \in V$. Edges, on the other hand, model the carrier-sensing relationships among links. There is an edge $e \in E$ between two vertices if the transmitters of the two associated links can sense each other. In this paper we will use the terms "links" and "vertices" interchangeably.

At any time, a link is in one of two possible states, active or idle. A link is active if there is a data transmission between its two end nodes. Thanks to carrier sensing, any two links that can hear each other will refrain from being active at the same time. A link sees the channel as idle if and only if none of its neighbors is active.

In ICN, each link maintains a *backoff timer*, $C$, the initial value of which is a random variable with an *arbitrary* distribution $f(t_{cd})$ and mean $E[t_{cd}]$. The timer value of the link decreases in a continuous manner with $dC/dt = -1$ as long as the link senses the channel as idle. If the channel is sensed busy (due to a neighbor transmitting), the countdown process is frozen and $dC/dt = 0$. When the channel becomes idle again, the countdown continues and $dC/dt = -1$ with $C$ initialized to the previous frozen value. When $C$ reaches 0, the link transmits a packet. The transmission duration is a random variable with *arbitrary* distribution $g(t_{tr})$ and mean $E[t_{tr}]$. After the transmission, the link resets $C$ to a new random value according to the distribution $f(t_{cd})$, and the process repeats. We define *the access intensity* of a link as the ratio of its mean transmission duration to its mean backoff time: $\rho = E[t_{tr}]/E[t_{cd}]$.

Let $s_i \in \{0,1\}$ denote the state of link $i$, where $s_i = 1$ if link $i$ is active (transmitting) and $s_i = 0$ if link $i$ is idle (actively counting down or frozen). The overall **system state** of ICN is $s = s_1 s_2 ... s_N$, where $N$ is the number of links in the network. Note that $s_i$ and $s_j$ cannot both be 1 at the same time if links $i$ and $j$ are neighbors because (i) they can sense each other; and (ii) the probability of them counting down to zero and transmitting together is 0 under ICN (because the backoff time is a continuous random variable).

The collection of feasible states corresponds to the collection of independent sets of the contention graph. An independent set (IS) of a graph is a subset of vertices such that no edge joins any two of them [8].

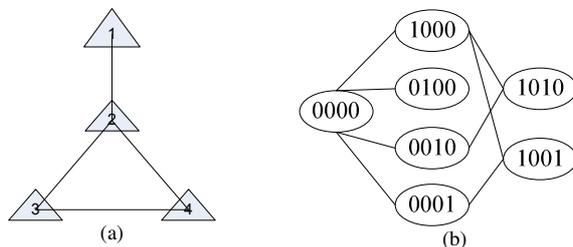

Fig. 1. (a) An example contention graph and (b) its state-transition diagram.

As an example, Fig. 1(b) shows the state-transition diagram of the contention graph in Fig. 1(a) under the ICN model. To avoid clutters, we have merged the two directional transitions between two states into one line in Fig. 1(b). Each transition from left to right corresponds to the beginning of the transmission of one particular link, while the reverse transition corresponds to the ending of the

[^1]: The mapping from network topology to contention graph has been studied in several prior works (e.g., [10]).

transmission of that link. For example, the transition $1000 \to 1010$ is due to link 3's beginning to transmit; the reverse transition $1010 \to 1000$ is due to link 3's completing its transmission.

*B. Equilibrium analysis*

If we further assume that the backoff time and transmission time are exponentially distributed, then $s(t)$ is a time-reversible Markov process. For any pair of neighbor states in the continuous-time Markov chain, the transition from the left state to the right state occurs at rate $1/E[t_{cd}]$, and the transition from the right state to the left state occurs at rate $1/E[t_{tr}]$.

Let $\mathcal{S}$ denote the set of all feasible states, and $n_s$ be the number of transmitting links when the system is in state $s = s_1 s_2 ... s_N$. The stationary distribution of state $s$ is given by [1] [2]:

$$P_s = \frac{\rho^{n_s}}{Z} \quad \forall s \in \mathcal{S}, \text{ where } Z = \sum_{s \in \mathcal{S}} \rho^{n_s} \quad (1)$$

The fraction of time during which link $i$ transmits is $th_i = \sum_{s:s_i=1} P_s$, which corresponds to the normalized throughput of link $i$.

Ref. [1] showed that (1) is in fact quite general and does not require the system state $s(t)$ to be a Markov process. In particular, (1) is insensitive to the distribution of the transmission duration $g(t_{tr})$, and the distribution of the backoff duration $f(t_{cd})$, given the ratio of their mean $\rho$.

Applying (1) to the state-transition diagram of Fig. 1 gives

$$P_{0000} = 1/Z = 1/(1 + 4\rho + 2\rho^2)$$
$$P_{1000} = P_{0100} = P_{0010} = P_{0001} = \rho/(1 + 4\rho + 2\rho^2) \quad (2)$$
$$P_{1010} = P_{1001} = \rho^2/(1 + 4\rho + 2\rho^2)$$

The normalized throughputs of the links are then given by

$$th_1 = P_{1000} + P_{1010} + P_{1001} = (\rho + 2\rho^2)/(1 + 4\rho + 2\rho^2)$$
$$th_2 = P_{0100} = \rho/(1 + 4\rho + 2\rho^2)$$
$$th_3 = P_{0010} + P_{1010} = (\rho + \rho^2)/(1 + 4\rho + 2\rho^2)$$
$$th_4 = P_{0001} + P_{1001} = (\rho + \rho^2)/(1 + 4\rho + 2\rho^2)$$

Note that $Z$ is a weighted sum of independent sets of $G$. In statistical physics, $Z$ is referred to as the partition function and the computation of $Z$ is the crux of many problems. We could also define $Z_i$ to be the weighted sum of the subset of independent sets in which $s_i = 1$. Then, $th_i$ could be equivalently expressed as $th_i = Z_i/Z$. This expression will be used later in the Appendix A.

---

[2] Here, we assume all links have the same access intensity $\rho = E[t_{tr}]/E[t_{cd}]$. For the case where different links have difference access intensities, (1) can be generalized by replacing $\rho^{n_s}$ with the $\prod_{i:s_i=1 \text{ in } s} \rho_i$.

## III. THROUGHPUT COMPUTATION USING BP

This section describes a direct application of BP in CSMA wireless networks: quick computation of the throughputs of links.

*A. Motivation*

Exact link-throughput computation requires the computation of $Z_i$ and $Z$, which is an NP-hard problem, since it involves finding all the independent sets of a contention graph. Thus, the problem can become intractable for large CSMA networks. As detailed in [1], for networks of more than 100 links, ICN computation can be rather time-consuming. An outstanding problem is to find quick and accurate approximate methods for large CSMA networks. This section is dedicated to this pursuit using BP.

*B. Graphical model in BP*

Under the framework of BP, the dependency between the states $s_i$ and $s_j$ of two neighbor vertices, $i$ and $j$, is captured using a compatibility function $\psi_{ij}(s_i, s_j)$, defined as follows:

$$\psi_{ij}(s_i, s_j) = \begin{cases} 0 & \text{if } s_i = 1, s_j = 1 \\ 1 & \text{otherwise} \end{cases} \quad (3)$$

In other words, the state $s_i = 1$ and the state $s_j = 1$ are not compatible because under CSMA, the two neighbor links cannot transmit together.

In addition, a weight is given to each possible state $s_i$ as follows:

$$\phi_i(s_i = 1) = \rho_i; \quad \phi_i(s_i = 0) = 1. \quad (4)$$

Note that $\phi_i(s_i = 1)$ and $\phi_i(s_i = 0)$ capture the relative likelihoods of states $s_i = 1$ and $s_i = 0$ if link $i$ were an isolated link without neighbors.

It is not difficult to verify that the stationary probability of the system state $s = s_1 s_2, ..., s_N$ in (1) can be rewritten as

$$P_s = \frac{\prod_{(i,j) \in E} \psi_{ij}(s_i, s_j) \prod_{i \in V} \phi_i(s_i)}{Z}, \forall s \in 2^N \quad (5)$$

In ICN, the normalized throughput of link $i$ is the marginal probability $p_i(s_i = 1) = \sum_{s:s_i=1} P_s$. In the context of BP, $p_i(s_i), s_i \in \{0,1\}$, corresponds to the belief at vertex $i$, denoted by $b_i(s_i)$.

*C. Message update rules in BP*

With the stationary distribution expressed in the form of (5), we next show how to use the BP algorithm to solve for $p_i(s_i = 1)$. The reader is referred to [8] for a general and detailed treatment of BP. Here, we focus on BP as applied to CSMA networks only.

When applying BP to CSMA networks, each vertex $i$ has an "intrinsic" belief of what the value of $p_i(s_i)$ should be. This intrinsic belief corresponds to the "on" probability when link $i$ is an isolated link. For an isolated link, $p_i(s_i = 1) = \rho_i/(1 + \rho_i) \propto \phi_i(s_i = 1)$, and $p_i(s_i = 0) =$

$1/(1+\rho_i) \propto \phi_i(s_i=0)$. That is, $p_i(s_i) \propto \phi_i(s_i)$ for an isolated link.

In addition, each vertex $i$ receives messages from its neighbors as to what they "think" $p_i(s_i)$ should be. Let $N_i$ denote the neighbors of vertex $i$ in $G$. Each neighbor $j \in N_i$ passes a message $m_{ji}(s_i)$ to $i$ as to its "belief" of $p_i(s_i)$. The beliefs of $i$ and all $j \in N_i$ are then aggregated into an overall belief in the form of a product:

$$b_i(s_i) = k_i \phi_i(s_i) \prod_{j \in N_i} m_{ji}(s_i), \quad (6)$$

where $k_i$ is a normalization constant so that $\sum_{s_i \in \{0,1\}} b_i(s_i) = 1$.

The messages are determined by the message update rule:

$$m_{ji}(s_i) \leftarrow \sum_{s_j \in \{0,1\}} \psi_{ij}(s_i, s_j) \phi_j(s_j) \prod_{k \in N_j \setminus i} m_{kj}(s_j)$$
$$= \sum_{s_j \in \{0,1\}} \psi_{ij}(s_i, s_j) b_j(s_j) / (k_j m_{ij}(s_j)) \quad (7)$$

Note that $\phi_j(s_j) \prod_{k \in N_j \setminus i} m_{kj}(s_j) \propto b_j(s_j)/m_{ij}(s_j)$. That is, it is proportional to the aggregated belief at vertex $j$ with the message from $i$ to $j$ factored out. In tree graphs, this message update rule can also be understood as the expression of the Bayes' formula [8].

The BP algorithm iterates (7) over all vertices $i$. In each iteration, we could normalize the messages according to $\sum_{s_i \in \{0,1\}} m_{ji}(s_i) = 1$ for $\forall j$ [3]. The iteration stops when $m_{ji}(s_i)$ converges.

It can be shown that (6) and (7) give exact solutions in tree graphs. Appendix A shows that in networks with a tree contention graph, the BP messages can be interpreted as the partition functions of subgraphs. This interpretation gives an explanation on why BP can give exact solutions in networks with loop-free graphs. Furthermore, each message needs only be computed once before convergence in trees. In other words, if there are no loops in the contention graph, (6) and (7) can solve ICN exactly within a time proportional to the number of edges in the graph. For loopy graphs, BP can often give good approximate results as well [8].

*D. Distributed BP*

BP can be easily implemented in a distributed manner. We focus on a particular vertex $j$. It stores a record of $N_j$ and the received messages from its neighbors, denoted by $M_j = \{m_{ij}(s_j), \forall i \in N_j\}$.

Each vertex $j$ operates as follows: Initially, vertex $j$ sets its outgoing messages $m_{ji}(s_i)$ to $\sum_{s_j \in \{0,1\}} \phi_j(s_j) \psi_{ij}(s_i, s_j)$, $\forall i \in N_j$. In each iteration, it passes $m_{ji}(s_i)$ to vertex $i$ and waits for time $T$ to receive messages from its neighbors. The locally stored messages in $M_j$ are then updated. Using the updated messages, vertex $j$ computes its outgoing messages according to (7) and repeats the iteration. The throughput of link $j$ (i.e., $th_j = b_j(s_j=1)$) can be computed based on the messages it stores according to (6). The pseudocode of distributed BP is given in Algorithm 1.

**Message Passing between Neighbors in $G$**

Distributed BP requires two neighbors who can mutually carrier-sense each other to exchange messages. Since the carrier-sensing range may be beyond the transmission range of regular data, the BP messages may need to be transmitted at a lower rate. The beacons in 802.11 typically use a lower data rate than the regular DATA packet, and BP messages can be carried on them. For further details, the reader is referred to [12], which proposed a scalable CSMA MAC protocol in which mutually interfering nodes also need to exchange information (note: look for the power-exchange algorithm in [12]).

**Periodical update to track dynamic network topology**

In practice, the network contention graph may change dynamically with new nodes joining and existing nodes leaving the network. Even among existing nodes, they may become idle when their users are not actively using the network. To track the variations of the network topology, $N_j$ needs to be refreshed periodically.

**Applications of distributed BP**

This distributed throughput computation algorithm provides an alternative way to estimate the throughputs of links in some network optimization problems. For example, in the adaptive CSMA algorithm in [7] and the "Wait-and-Hop" link frequency assignment algorithm in [13], decisions in each iteration are made based on the throughputs of links under the link access intensities and link frequency assignments of the last iteration, respectively. Both papers proposed to use real-time measurements to gather the throughputs of links. Accurate real-time throughput measurements, however, take time, especially in networks susceptible to temporal starvation [6]. In such a network, the throughput of a link can alternate between 0 and 1 in cycles of very long duration. To avoid triggering oscillations in the control mechanism, each measurement must be averaged over several such cycles. As a result, the optimization algorithms may converge slowly.

In contrast, the throughput computation using BP does not have such problems. The speed of convergence is determined by how frequently the links pass BP messages to each other. Our simulation in Section III-*F* shows that for a network of up to 200 links, BP converges within 100 iterations. In real networks (e.g., WLAN), we may use beacons to exchange BP messages. Each AP typically broadcasts a beacon every 0.1 second [9]; thus, distributed BP can give solutions within ten seconds. If BP messages are piggybacked onto the regular DATA packets, the speed of convergence can be even faster.

| Algorithm 1: Distributed BP |
|---|

1. The following procedure runs on each individual vertex independently. We focus on a particular vertex $j$.

---
[3] If we normalize the beliefs in (6) without normalizing the messages, the magnitudes of the messages may grow unbounded, but not the beliefs themselves. Thus, the algorithm may still be well-behaved if the beliefs converge quickly.

2. Vertex $j$ keeps track of its one-hop neighbors $N_j$ and the incoming messages $M_j = \{m_{ij}(s_j), \forall i \in N_j\}$.
3. In distributed BP, $N_j$ are periodically refreshed.

4. **procedure** INITIALIZATION
5.   $m_{ji}(s_i)$, $\forall i \in N_j$ $\leftarrow$ $\sum_{s_j \in \{0,1\}} \phi_j(s_j) \psi_{ij}(s_i, s_j)$
6. **end procedure**

7. **procedure** ITERATION
8.   Pass $m_{ji}(s_i)$ to vertex $i$ $\forall i \in N_j$
9.   Wait for time $T$ to receive messages from its neighbors, $m_{ij}(s_j)$, $\forall i \in N_j$ and update $M_j$ accordingly.
10.  Compute $m_{ji}(s_i)$, $\forall i \in N_j$ according to (7)
11.  Invoke procedure BELIEFCOMPUTATION and repeat procedure ITERATION
12. **end procedure**

13. **procedure** BELIEFCOMPUTATION
14.  Compute its belief $b_j(s_j)$ according to (6) and in turn obtain throughput $th_j$.
15. **end procedure**

*E. BP in loopy contention graphs*

Although BP can often give good approximations, as pointed out in [14], if we apply BP in loopy contention graphs, the information may circulate indefinitely around the loops, and BP may give inaccurate solutions and even not converge.

Consider a triangular graph consisting of three vertices. In BP, messages are passed between each pair of neighboring vertices. Vertex 1 gives certain information to vertex 2, some of which is included in the information from vertex 2 to vertex 3 and finally passed back to vertex 1, where it is regarded as a "new" incoming message. This message contains information correlated with the original information at vertex 1. The message update rule and the belief computation formula, however, do not take this correlation into account.

When the loop is large, the information a vertex $i$ gives out vanishes along the cycle back to $i$, resulting in a smaller computation error. It can be shown that BP converges to the fixed point $b_i(s_i = 0) = (1 + \sqrt{1+4\rho})/2\sqrt{1+4\rho}$ for each vertex in any $N$-vertex ring graph regardless of $N$ (See Appendix D). Given a value of $\rho = \rho_0 = 83/15.5$ (typical in 802.11 networks), the errors of BP for different $N$ are
- 8% for the 3-vertex ring;
- 0.1% for the 8-vertex ring;
- Zero for the $N$-vertex ring as $N \to \infty$.

That is, the error of BP decreases as the length of the cycle increases.

From the ring example, we can see intuitively that for general graphs, small loops are the loops that cause the more significant errors. To contain the errors, we want to eliminate small loops in message propagation. This is the basic idea behind the generalized belief propagation (GBP). For easy comparison, in the next subsection we evaluate the performance of GBP together with BP first, leaving the theoretical details of GBP to Section VI.

*F. Experimental Evaluation*

Ref. [1] proposed a quick "back-of-the-envelope" (BoE) algorithm for link throughputs computation in CSMA wireless networks. BoE could only handle networks of up to 50 links. Networks of larger size were left as an open issue. The focus of our experiment here is on the accuracy and speed of BP and GBP for networks of more than 50 links.

We implement both algorithms in a centralized manner using MATLAB programs. The simulations run on an IBM ThinkCentre M51 Desktop computer with 3.4GHz Intel Pentium 4 processor. The throughputs computed by BP and GBP are compared with that obtained from an ICN-simulator to examine their accuracy. The CPU runtimes are presented to evaluate the speed of BP and GBP. Furthermore, we list the average number of iterations a link performs before convergence. This will be used to estimate the convergence time for distributed implementation in real networks. In our experiments, we define the minimum $n$ such that $\max_j |th_j[n] - th_j^*| / th_j^* < 1\%$ is satisfied as the number of iterations for BP and GBP to achieve convergence, where $th_j^*$ is the final converged value[4].

In the first set of experiments, we randomly generate networks of different numbers of links. We vary the network area while maintaining the mean degree of links (number of neighbors per link) to around four. The access intensities of all links are set to $\rho = \rho_0 = 83/15.5$, which corresponds to that typically seen in 802.11b networks. For each link, we calculate the error of the throughput obtained by BP and GBP relative to the simulated throughput obtained from the ICN simulator. The error is normalized by the maximum link throughput in the network. For each network setting, we randomly generate ten different topologies and the experimental data are averaged over the ten networks.

As shown in Table I, for networks of up to 200 links, the error of BP is kept to 7.0% or below, while the error of GBP is consistently lower than 1%. The maximum error of GBP is about 0.6%. As for computation complexity, BP is very fast while GBP can also output solutions within seconds.

TABLE I. MEAN LINK THROUGHPUT ERRORS, RUNTIMES AND NUMBERS OF ITERATIONS OF BP AND GBP FOR NETWORKS IN WHICH EACH VERTEX HAS ON AVERAGE FOUR NEIGHBORS.

| # of links | | 50 | 100 | 200 |
|---|---|---|---|---|
| Accuracy | BP | 5.3% | 6.5% | 7.0% |
| | GBP | 0.3% | 0.3% | 0.6% |
| Speed (runtime in second) | BP | 0.02 | 0.06 | 0.23 |
| | GBP | 1.56 | 2.53 | 6.84 |
| Number of Iterations | BP | 14 | 15 | 15 |
| | GBP | 22 | 28 | 32 |

Table II shows the scenario in which the number of links is fixed to 100 while the network area is varied. That is, the mean degree of links is varied. Again, GBP gives more accurate results while costing more CPU time. The error of BP is still below 10%.

---
[4] We use exponential averaging to smooth out the computed messages for GBP algorithms: i.e., $m_{ave}[n] = (1-\alpha) m_{ave}[n] + \alpha m[n]$, $0 < \alpha < 1$, where $m[n]$ is the newly computed message and $\alpha$ is the smoothing factor. $m[n]$ is recomputed in each iteration. For BP algorithms, we do not perform the procedure above because it converges smoothly even without the procedure.

TABLE II. MEAN LINK THROUGHPUT ERRORS, RUNTIMES AND NUMBERS OF ITERATIONS OF BP AND GBP FOR NETWORKS OF 100 LINKS.

| Mean Vertex Degree | | 2 | 4 | 6 |
|---|---|---|---|---|
| Accuracy | BP | 4.7% | 7.0% | 7.2% |
| | GBP | 0.2% | 0.3% | 0.3% |
| Speed (runtime in second) | BP | 0.02 | 0.06 | 0.11 |
| | GBP | 1.33 | 2.53 | 14.34 |
| Number of Iterations | BP | 13 | 15 | 16 |
| | GBP | 24 | 28 | 35 |

In Table I and Table II, $\rho$ is set to $\rho_0 = 83/15.5$. A question is how well these algorithms work under different $\rho$. It is known that when $\rho$ is large, two neighbor vertices become more tightly coupled, and the message passing within a loop may incur more computational errors. Table III shows the accuracy of both algorithms for different $\rho$ in a 100-link network with the mean degree of links equal to four. As can be seen, the mean error of BP increases with the value of $\rho$. More impressive is GBP, whose mean error is very small even for $\rho = 4\rho_0$. This shows that GBP performs well over a large range of $\rho$.

TABLE III. MEAN LINK THROUGHPUT ERRORS, RUNTIMES AND NUMBERS OF ITERATIONS OF BP AND GBP FOR NETWORKS OF DIFFERENT $\rho$.

| $\rho / \rho_0$ | | 2 | 3 | 4 |
|---|---|---|---|---|
| Accuracy | BP | 10.6% | 12.3% | 13.5% |
| | GBP | 0.2% | 0.3% | 0.3% |
| Speed (runtime in second) | BP | 0.08 | 0.09 | 0.11 |
| | GBP | 4.55 | 4.84 | 5.03 |
| Number of Iterations | BP | 32 | 57 | 76 |
| | GBP | 45 | 78 | 92 |

For all the scenarios, both algorithms converge within dozens of iterations. That is, if implemented in a distributed manner in which each link passes a message every 0.1 second (e.g., we use beacons for message passing in a 802.11 network), both BP and GBP can obtain links throughputs within seconds in real networks.

## IV. COMPUTATION OF LINK ACCESS INTENSITIES GIVEN TARGET LINK THROUGHPUTS

This section proposes an inverse belief propagation (IBP) algorithm to compute the link access intensities required to meet target link throughputs. We show that IBP can be easily implemented in a distributed manner and only only-hop message passing is needed. We evaluate the speeds and accuracies of IBP and IGBP (to be presented in Section VI) by simulations.

### A. Motivation

In network design, an interesting problem is as follows. Given a network contention graph $G$ and a set of target link throughputs, how to set the link access intensities $\bar{\rho}$ to meet the target link throughputs.

For small networks, we can find $\bar{\rho}$ by solving (1) and $th_i = \sum_{s:s_i=1} P_s$. However, similar to the throughput computation using (1), the computation becomes intractable when the network is large. IBP below gives appropriate approximate solutions within a short time.

### B. Definition of IBP

As described in Section III, the operation of BP is as follows. Given the contention graph of the network $G$ and the access intensities of links $\bar{\rho}$, BP computes the throughputs of links. That is, $\bar{th} = BP(G, \bar{\rho})$.

**Definition of IBP:** We define $\bar{\rho} = BP^{-1}(G, \bar{th})$ as the inverse operation of belief propagation for $\bar{th} = BP(G, \bar{\rho})$, where $\bar{th}$ is the vector of target link throughputs.

### C. Message update rules and its distributed implementation

#### 1) Message update rules in IBP

As mentioned in Section III-B, the belief at vertex $j$ $b_j(s_j = 1)$ corresponds to the link throughput. That is, the belief of each link $j$, $b_j(s_j)$ is given in IBP.

From (7) we obtain the message update rule

$$m_{ji}(s_i) \leftarrow \sum_{s_j \in \{0,1\}} \psi_{ij}(s_i, s_j) b_j(s_j) / (k_j m_{ij}(s_j)) \quad (8)$$

and from (6) we have

$$\rho_j = \phi_j(s_j = 1) = \frac{b_j(s_j = 1) \prod_{i \in N_j} m_{ij}(s_j = 0)}{b_j(s_j = 0) \prod_{i \in N_j} m_{ij}(s_j = 1)} \quad (9)$$

The IBP algorithm iterates (8) over all vertices $j$. Similar to BP, in each iteration we could normalize the messages according to $\sum_{s_i \in \{0,1\}} m_{ji}(s_i) = 1, \forall j$. The iteration stops when $m_{ji}(s_i)$ converges or a maximum number of iterations is reached.

Note that IBP, being an approximate algorithm, has computation errors which potentially can result in non-convergence of the algorithm. As will be demonstrated in Section VI-D, we can resort to IGBP for more accurate computation. Another reason for non-convergence is due to the problem formulation itself. We require the target $\bar{th}$ to be feasible and then seek the $\bar{\rho}$ to achieve that. If the given $\bar{th}$ is beyond the feasible region (as defined in Section II-C of [7]), then no matter what algorithm we use, there is no solution. Formulating the problem as a system utility optimization problem as in Section V removes this difficulty, as the algorithm would then iterate to zoom into a feasible $\bar{th}$ that can achieve optimal system utility.

#### 2) Distributed IBP

In real applications, it is desirable to make IBP work in a distributed manner. Again we focus on a particular vertex $j$

that knows its target throughput $th_i$ (i.e., $b_j(s_j)$). Similar to distributed BP, vertex $j$ stores a record of $N_j$ and the messages from its neighbors, $M_j = \{m_{ij}(s_j), \forall i \in N_j\}$.

The procedure that link $j$ operates is as follows: Initially, vertex $j$ sets its outgoing messages $m_{ji}(s_i)$ to $\sum_{s_j \in \{0,1\}} \psi_{ij}(s_i, s_j) b_j(s_j)$, $\forall i \in N_j$. In each iteration, it passes $m_{ji}(s_i)$ to vertex $i$ and waits for time $T$ to receive messages from its neighbors. The locally stored messages in $M_j$ are updated accordingly. Using the updated messages, it computes its outgoing messages according to (8) and repeats the iteration. The access intensity $\rho_j$ is computed according to (9). The pseudocode of distributed IBP is largely similar to that of distributed BP. Here, we only show the parts that are different.

---

**Algorithm 2: Distributed IBP**

5. $m_{ji}(s_i)$, $\forall i \in N_j$ ← $\sum_{s_j \in \{0,1\}} \psi_{ij}(s_i, s_j)$
10.     Compute $m_{ji}(s_i)$, $\forall i \in N_j$ according to (8)
11.     Invoke procedure ACCESSINTENSITYCOMPUTATION and repeat procedure ITERATION;
12. **end procedure**

13. **procedure** ACCESSINTENSITYCOMPUTATION
14.     Compute its access intensity $\rho_j$ according to (9)
15. **end procedure**

---

*3) Convergence of IBP*

With respect to the convergence of IBP, we have the following theorem:

**Theorem 1:** If the target throughput is feasible in the sense that $\overline{th} = BP(G, \overline{\rho})$ for some $\overline{\rho}$, IBP defined by (8) and (9) is a contraction mapping and is guaranteed to converge to $\overline{\rho}$.

**Proof:** See Appendix B.

We have shown that IBP is guaranteed to converge. However, recall that $\overline{th} = BP(G, \overline{\rho})$ is an approximation of the actual link throughputs in the CSMA network. Similarly, IBP may output a $\overline{\rho}$ that does not exactly yield the target $\overline{th}$ in the actual network.

To reduce the errors in loopy graphs, we can also adapt GBP for the reverse operation. The details of IGBP will be presented in Section VI-D.

### D. Experimental Evaluation

We examine the performance of IBP and IGBP. First, consider Network 1 shown in Fig. 2. Define $0 \leq \gamma < 1$ as the "load factor". The target throughput vector is set to
$\overline{th} = \gamma * [0.2*(1,0,1,0,0,0) + 0.3*(1,0,0,1,0,1) + 0.2*(0,1,0,0,1,0) + 0.3*(0,0,1,0,1,0)] = \gamma * (0.5, 0.2, 0.5, 0.3, 0.5, 0.3)$.
That is, we set $\overline{th}$ to be a linear combination of some MaIS, multiplied by a factor $\gamma < 1$ to make sure that the target throughput vector is within the capacity region as in [7].

We implement IBP and IGBP using MATLAB programs. For Network 1, we vary $\gamma$ and find the corresponding access intensities $\overline{\rho}$ to meet the link target throughputs using IBP (IGBP). We then use an ICN-simulator to get the throughputs of Network 1 with the access intensities $\overline{\rho}$ found. For each link, we calculate the error of the throughput obtained by the ICN-simulator relative to the target throughput. The error is normalized by the maximum link throughput in the network. In our experiments, we define the minimum $n$ such that $\max_j |\rho_j[n] - \rho_j^*| / \rho_j^* < 1\%$ is satisfied as the number of iterations for IBP and IGBP to achieve convergence, where $\rho_j^*$ is the final converged value.

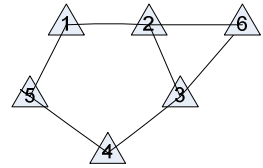

Fig.2. Network 1.

Table IV shows the mean throughput errors of IBP and IGBP with respect to $\gamma$. As can be seen, when $\gamma$ is not large (e.g., below 0.6), both IBP and IGBP work quite well. As $\gamma$ approaches 1, IBP has a throughput error of 12.3% and IGBP has an error of 3.8%. It is known that as $\gamma$ increases, we need larger access intensities to meet the target throughputs. As shown in Section III, the mean error of BP and GBP increases with the value of $\rho$. Based on the same framework, the errors of IBP and IGBP will also increase with the value of $\rho$. This explains why the errors of IBP and IGBP increase with $\gamma$. As for computation complexity, IBP is very fast and its runtime is very close to zero while IGBP can output solutions within one second.

Next we conduct a set of random graph experiments as follows. We randomly generate networks of different numbers of links. The mean vertex degree is four. The access intensity of link $i$, $\rho_i$, is randomly generated within the interval $[\rho_0, 4\rho_0]$. Then we run the ICN simulator to get the link throughputs and set them to be the target throughputs of IBP and IGBP. Finally we run IBP and IGBP and examine the throughput errors of both algorithms. Table V shows the mean throughput errors, CPU runtimes and number of iterations of IBP and IGBP. For networks of up to 200 links, the error of IBP is kept to 6.2% or below, while the error of IGBP is within 1%. As for computation complexity, IBP is very fast while IGBP can also give solutions within seconds.

Table VI shows the scenario in which the number of links is fixed to 100 while the mean degree of links is varied. Again, the throughput error of IBP is kept to below 4%. IGBP's throughput error is below 1%. As for computation complexity,

IBP is very fast, while IGBP is slower but still outputs solutions within a minute. Note that these CPU runtimes are from a centralized implementation in which there is only one processor computing the results. Distributed implementation will be much more scalable with the number of links. As can be seen, both IBP and IGBP converge within dozens of iterations. If beacons in real networks are used for message passing, IBP and IGBP can output solutions within seconds for network of up to 200 links.

TABLE IV. MEAN THROUGHPUT ERRORS, RUNTIMES AND NUMBER OF ITERATIONS OF IBP AND IGBP FOR NETWORK 1.

| $\gamma$ | | 0.2 | 0.4 | 0.6 | 0.8 | 0.9 | 0.98 |
|---|---|---|---|---|---|---|---|
| Accuracy | IBP | 0.4% | 0.8% | 2.2% | 6.1% | 9.2% | 12.3% |
| | IGBP | 0.1% | 0.2% | 0.4% | 1.7% | 2.7% | 3.8% |
| Speed (runtime in second) | IBP | 0 | 0 | 0 | 0 | 0 | 0 |
| | IGBP | 0.27 | 0.28 | 0.29 | 0.31 | 0.32 | 0.39 |
| Number of Iterations | IBP | 2 | 2 | 3 | 6 | 12 | 18 |
| | IGBP | 15 | 20 | 23 | 26 | 30 | 37 |

TABLE V. MEAN THROUGHPUT ERRORS, RUNTIMES AND NUMBERS OF ITERATIONS OF IBP AND IGBP FOR NETWORKS WITH CONTENTION GRAPHS IN WHICH EACH VERTEX HAS ON AVERAGE FOUR NEIGHBORS.

| # of links | | 50 | 100 | 200 |
|---|---|---|---|---|
| Accuracy | IBP | 5.40% | 3.27% | 6.2% |
| | IGBP | 0.11% | 0.07% | 0.79% |
| Speed (runtime in second) | IBP | 0.02 | 0.03 | 0.07 |
| | IGBP | 2.67 | 9.68 | 10.73 |
| Number of Iterations | IBP | 29 | 38 | 47 |
| | IGBP | 42 | 44 | 54 |

TABLE VI. MEAN THROUGHPUT ERRORS, RUNTIMES AND NUMBERS OF ITERATIONS OF IBP AND IGBP FOR NETWORKS OF 100 LINKS.

| Mean vertex degree | | 2 | 4 | 6 |
|---|---|---|---|---|
| Accuracy | IBP | 2.56% | 3.27% | 3.69% |
| | IGBP | 0.02% | 0.07% | 0.77% |
| Speed (runtime in second) | IBP | 0.02 | 0.03 | 0.12 |
| | IGBP | 2.57 | 9.68 | 40.71 |
| Number of Iterations | IBP | 2 | 38 | 58 |
| | IGBP | 11 | 44 | 74 |

## V. BP-ADPATIVE CSMA (BP-ACSMA)

This section investigates solving the network utility optimization problem in CSMA networks using BP.

### A. Motivation and Problem Formulation

In the previous section, the target link throughputs are given, and the corresponding link access intensities are computed. A problem is that in general we do not know whether the target link throughputs are feasible or not — computation of the feasible region is itself a tough problem for large networks. A way to circumvent this problem is to focus on optimizing a system utility $\sum_j U_j(th_j)$ instead, where $U_j(th_j)$ is the utility of link $j$. That is, we aim to find $\bar{\rho}$ to optimize $\sum_j U_j(th_j)$. In the following, we briefly review the background leading to the ACSMA. Then, in Part C, we introduce the alternative of using BP to solve the problem.

Recall that the feasible states of ICN are the independent sets of the contention graph. Define an indicator function $x_j^s$ such that $x_j^s = 1$ if link $j$ is transmitting in state $s$ and $x_j^s = 0$ otherwise. Let $u_s$ be the probability of state $s$ (i.e., fraction of airtime dedicated to state $s$). Furthermore, let $f_j$ denote the input rate of link $j$. Let $\bar{u}$ and $\bar{f}$ denote the vectors consisting of $u_s$ for all $s$, and $f_j$ for all $j$, respectively. Consider the following utility optimization problem:

$$\max_{\bar{u},\bar{f}} \sum_j U_j(f_j)$$
$$\text{s.t.} \quad \sum_s u_s x_j^s \geq f_j \quad \forall j \qquad (10)$$
$$u_s \geq 0 \ \forall s ; \ \sum_s u_s = 1$$

As explained in [7], when the system is in a state $s$ and $x_j^s = 1$, but link $j$ has no packet in its queue, link $j$ will transmit a dummy packet. This accounts for the inequality $\sum_s u_s x_j^s \geq f_j$ in order to balance the input and output traffic.

The optimization problem as formulated in (10) has two problems. First, it is a difficult combinatorial optimization problem. Also, it is not clear how to implement a distributed algorithm to solve it. Second, even if a solution could be found, to realize it using CSMA, the $\bar{u}$ found would still have to be mapped to $\bar{\rho}$. That is, $u_s$ must be equal to the stationary probability $P_s = \prod_{j:x_j^s=1} \rho_j / Z$ for CSMA networks.

To circumvent the above difficulties, [7] formulated an alternative optimization problem as follows:

$$\max_{\bar{u},\bar{f}} \beta \sum_j U_j(f_j) - \sum_s u_s \log u_s$$
$$\text{s.t.} \quad \sum_s u_s x_j^s \geq f_j \quad \forall j \qquad (11)$$
$$u_s \geq 0 \ \forall s ; \ \sum_s u_s = 1$$

Compared with (10), the objective function in (11) has an extra entropy term $-\sum_s u_s \log u_s$. When $\beta$ is large, (11) asymptotically approaches (10). As shown below, the $\bar{u}$ found by (11) turns out to be CSMA realizable. Indeed $r_j = \log(\rho_j)$ turns out to be the dual variable to the constraint $\sum_s u_s x_j^s \geq f_j$.

Associate dual variable $r_j$ with the constraint $\sum_s u_s x_j^s \geq f_j \quad \forall j$, without assuming $r_j = \log(\rho_j)$ for the time being. A partial Lagrangian of problem (11) is

$$L(\bar{u},\bar{r},\bar{f}) = \beta \sum_j U_j(f_j) - \sum_s u_s \log u_s + \sum_j r_j \left( \sum_s u_s x_j^s - f_j \right) \quad (12)$$

Given $\bar{r}$ and $\bar{f}$, the optimal $\bar{u}$ to (12) can be shown to be

$$u_s^*(\bar{r}) = \exp\left(\sum_j r_j x_j^s\right) \Big/ \sum_s \exp\left(\sum_j r_j x_j^s\right), \forall s \quad (13)$$

We see that (13) is just the stationary distribution of CSMA networks with $r_j = \log(\rho_j) \quad \forall j$.

The optimal $\bar{f}$ is given by

$$\partial L(\bar{u},\bar{r},\bar{f}) \big/ \partial f_j = \beta U_j'(f_j) - r_j = 0 \quad (14)$$

The optimal $\bar{r}$ is given by
$$\partial L(\bar{u},\bar{r},\bar{f})/\partial r_j = \sum_s u_s x_j^s - f_j = 0 \quad (15)$$
Combining (13), (14) and (15), we find that the optimal solution to (11) is given by a set of $\bar{r}$ and $\bar{f}$ that satisfy
$$\begin{aligned}\beta U'_j(f_j) - \log(\rho_j) &= 0 \\ th_i = f_i &= \sum_s u_s x_j^s\end{aligned} \quad (16)$$

In Section B below, we briefly review how [7] solves the optimization problem using a distributed adaptive CSMA algorithm. We present an alternative method using the BP framework in Section C.

### B. ACSMA proposed in [7]

The joint scheduling and congestion control algorithm (ACSMA) proposed in [7] looks for the optimal solution to (11) by steepest ascent of $L(\bar{u},\bar{r},\bar{f})$. According to (15), $\partial L(\bar{u},\bar{r},\bar{f})/\partial r_j =$ output rate of link $j$ − input rate of link $j$. The queue size of link $j$ is a smoothed measure of the difference in the output rate and input rate. Thus, in each iteration, link $j$ adjusts its $\rho_j$ such that $r_i = \log(\rho_i)$ is proportional to its queue length. If the input rate of the queue is larger than the service rate, the queue builds up, leading to an increase in $\rho_j$, and vice versa. Note, that $\rho_j$ controls the output rate of link $j$. For the input rate $f_j$, link $j$ adjusts $f_j$ to satisfy $\beta U'_j(f_j) - \log(\rho_j) = 0$ in (14) based on the newly computed $\rho_j$. Before the next iterative update, link $j$ waits for some time to examine whether the load $f_j$ can be supported by the network under current $\bar{\rho}$ through its queue size. The iterations continue until the overall network finds a set of access intensities $\bar{\rho}$ that can support the loads $\bar{f}$. At that point, the throughput of link $j$ satisfies $th_j = f_j$.

ACSMA does not explicitly "compute" the link throughputs using (13). Rather, it makes use of actual data packets to probe the network and "measure" the link throughputs. To smooth out the measurement due to temporal throughput fluctuations to which CSMA networks are susceptible, long smoothing interval between successive iterations may be required.

### C. BP-ACSMA

The optimal network utility in (11) is achieved when the link access intensities and throughputs are such that (16) holds. BP can be applied to make sure that (16) is satisfied.

*1) Message update rules of BP-CSMA*

In BP-ACSMA, the messages are determined by the message update rule:
$$m_{ji}(s_i) \leftarrow \sum_{s_j \in \{0,1\}} \psi_{ij}(s_i,s_j)\phi_j(s_j)\prod_{k \in N_j \setminus i} m_{kj}(s_j) \quad (17)$$

In each iteration, based on the received messages, vertex $j$ computes belief $b_j(s_j)$ according to
$$b_j(s_j) = k_j\phi_j(s_j)\prod_{i \in N_j} m_{ij}(s_j) \quad (18)$$

It then solves for $\rho_j$ from (16) by setting $th_j = b_j(s_j = 1)$. Based on the new $\rho_j$, vertex $j$ updates messages $m_{ji}(s_i)$, $i \in N_j$ according to (17) and broadcasts the messages to its neighbors.

In essence, BP replaces the network probing and throughput measurement in ACSMA by computation.

*2) Distributed implementation*

Consider a particular vertex $j$. It locally stores a record of $N_j$, the messages from its neighbors $M_j = \{m_{ij}(s_j), \forall i \in N_j\}$, and its utility function $U_j(th_j)$. $N_j$ is periodically refreshed to track the dynamics of the local network contention graph.

Initially, vertex $j$ sets its outgoing messages $m_{ji}(s_i)$ to $\sum_{s_j \in \{0,1\}} \psi_{ij}(s_i,s_j)$, $\forall i \in N_j$. In each iteration, it passes $m_{ji}(s_i)$ to each neighbor vertex $i$. It then waits for time $T$ to receive messages from its neighbors. Based on the received messages, link $j$ then (i) computes its belief $b_j(s_j)$ using (18); (ii) solves for $\rho_j$ according to (16); and (iii) determines its outgoing messages according to (17) using the newly computed $\rho_j$ in (ii).

The pseudocode of BP-ACSMA is largely similar to that of distributed BP. Here, we only show the parts that are different.

---
**Algorithm 3: BP-ACSMA**

5. $m_{ji}(s_i)$, $\forall i \in N_j$ ← $\sum_{s_j \in \{0,1\}} \psi_{ij}(s_i,s_j)$
10. Invoke procedure ACCESSINTENSITYCOMPUTATION and repeat procedure ITERATION;
11. **end procedure**

12. **procedure** ACCESSINTENSITYCOMPUTATION
13.     Compute its belief $b_j(s_j)$ using (18);
14.     Solve for $\rho_j$ according to (16)
15.     Compute $m_{ji}(s_i)$, $\forall i \in N_j$ according to (17)
16. **end procedure**

---

### D. Experimental Evaluation

We evaluate the performance of both BP-ACSMA and GBP-ACSMA (details of GBP-ACSMA will be presented in Section VI-E). We consider the proportional fairness utility: $U_j(th_j) = \log(th_j)$, and set the weighting factor $\beta$ to 1. We implement both algorithms using MATLAB programs. For both algorithms, the outputs are the converged link access intensities, $\bar{\rho}^*$. To evaluate the performance of each algorithm, we use the ICN simulator to get the throughputs of networks under the $\bar{\rho}^*$ found, and then obtain the network utility achieved from the throughputs. Recall that $\rho_j$ relates to $r_j$ in ACSMA of [7] via $r_j = \log(\rho_j)$. For easy comparison between our BP-based algorithms and ACSMA, we use $r_j$ in our convergence test although the parameters being adjusted in our algorithms are $\rho_j \ \forall j$. Let $r_j[n]$ be the value of $r_j$ in iteration $n$. We define the number of iterations required for convergence in BP-ACSMA (GBP-ACSMA) as the minimum $n$ such that $\max_j |r_j[n] - r_j^*|/r_j^* < 1\%$, where $r_j^* = \log(\rho_j^*)$ is the final converged $r_j$ value.

We also implement ACSMA of [7]. In ACSMA, the parameters adjusted in iteration $n$ are $r_j[n]$ and $f_j[n] \ \forall j$. If ACSMA converges, then $r_j[n]$ and $f_j[n]$ will asymptotically approach the targeted $r_j^*$ and $f_j^*$ as $n$ increases. We define the minimum $n$ such that $\max_j |r_j[n] - r_j^*|/r_j^* < 3\%$ is satisfied as the number of iterations for ACSMA to achieve convergence. Note that here we use a looser convergence test for ACSMA; by nature, some fluctuations are unavoidable in ACSMA even after convergence because of its measurement approach. In our simulation, the update interval of ACSMA is set to 150 DATA packet times to guarantee that convergence can be achieved. We find that if the update interval is set to 125 DATA packet times, ACSMA cannot converge in some networks we test[5].

In the first set of experiments, we randomly generate networks with different numbers of links. The mean degree of links is around four. In each simulation run, we gather the statistics of two metrics: i) normalized total system throughput $Th = \sum_j th_j$; ii) system utility $U = \sum_j \log(th_j)$. Table VII shows the achieved throughputs and network utilities of BP-ACSMA, GBP-ACSMA and ACSMA. As shown, BP-ACSMA has acceptable performance in terms of both throughputs and network utilities; and GBP-ACSMA has comparable performance to ACSMA. As for speed, BP-ACSMA and GBP-ACSMA output solutions after dozens of iterations while ACSMA often requires hundreds of iterations.

TABLE VII. ACHIEVED AGGREGATE THROUGHPUTS, UTILITIES AND NUMBER OF ITERATIONS OF BP-ACSMA, GBP-ACSMA AND ACSMA FOR NETWORKS WITH CONTENTION GRAPHS IN WHICH EACH VERTEX HAS ON AVERAGE FOUR NEIGHBORS

| # of links | | 25 | 50 | 75 |
|---|---|---|---|---|
| BP-ACSMA | Th | 8.63 | 17.16 | 26.28 |
| | U | -34.24 | -68.97 | -95.75 |
| | Number of Iterations | 7 | 9 | 9 |
| GBP-ACSMA | Th | 8.38 | 16.25 | 25.26 |
| | U | -31.85 | -65.41 | -90.76 |
| | Number of Iterations | 34 | 43 | 44 |
| ACSMA | Th | 8.12 | 15.94 | 24.74 |
| | U | -30.58 | -62.47 | -89.14 |
| | Number of Iterations | 296 | 311 | 382 |

[5] This brings up another issue with ACSMA. That is, we do not know how to set the update interval $T$ in an optimal manner beforehand, and we need to run the algorithm to determine the minimum $T$ required for each network. BP-ACSMA and GBP-CSMA, however, do not have this issue because the update interval is not related to measurement smoothing time needed.

In the second set of experiments, we randomly generate networks of 100 links with varying mean vertex degrees. Table VIII compares the three algorithms. As the network becomes denser, more loops appear in the contention graph, resulting in more computation error of BP-ACSMA. As shown in Table VIII, BP-ACSMA loses accuracy when the mean vertex degree is set to six. GBP-ACSMA continues to work well since it has removed loops in message passing (see Section VI-E). Table VIII also shows that BP-ACSMA and GBP-ACSMA achieve higher aggregate throughputs than ACSMA does with some utility loss. As for convergence speed, BP-ACSMA and GBP-ACSMA are much faster than ACSMA.

TABLE VIII. ACHIEVED AGGREGATE THROUGHPUTS, UTILITIES AND NUMBER OF ITERATIONS OF BP-ACSMA, GBP-ACSMA AND ACSMA FOR NETWORKS OF 100 LINKS.

| Mean Vertex Degree | | 2 | 4 | 6 |
|---|---|---|---|---|
| BP-ACSMA | Th | 43.89 | 32.74 | 26.55 |
| | U | -92.53 | -152.64 | -218.03 |
| | Number of Iterations | 9 | 9 | 11 |
| GBP-ACSMA | Th | 46.33 | 34.45 | 25.79 |
| | U | -90.64 | -139.70 | -179.65 |
| | Number of Iterations | 39 | 52 | 64 |
| ACSMA | Th | 43.49 | 32.14 | 25.12 |
| | U | -90.45 | -121.31 | -147.91 |
| | Number of Iterations | 228 | 405 | 413 |

### E. Comparison of BP-ACSMA and GBP-ACSMA

Our simulations in Part D show that both BP-ACSMA and GBP-ACSMA converge within dozens of iterations for a network of 100 links. ACSMA converges only after hundreds of iterations. For comparison, let us map the number of iterations to time needed for convergence in real network operation.

For BP-ACSMA and GBP-ACSMA, beacons could be used for message passing. In 802.11 networks, typically a beacon is broadcasted every 0.1s. For BP-ACSMA, from the results in Tables VII and VIII, convergence is achieved within 11 iterations for all the scenarios we tested. Using beacons for message passing (note that the transmission time of a beacon is about 0.1ms. Thus, each time a link has sufficient time to broadcast its message between two successive iterations), it only needs 0.1*11 =1.1s to output solutions for networks of up to 100 links. For GBP-ASMA, convergence is achieved within 64 iterations for all the scenarios tested, corresponding to a convergence time of within 6.4 seconds. The convergence speed of both algorithms can be even faster if the messages are piggybacked on data packets rather than being carried on beacons. By contrast, ACSMA requires 413*150ms ≈ 62s for convergence, assuming a DATA packet duration is 1ms — recall that we experimentally found that we need 150 DATA packet times for convergence of ACSMA in the networks simulated.

For networks that exhibit temporal starvation, even more time is needed for ACSMA for each iteration to smooth out the measurement. An example of a network that exhibits temporal starvation is Cayley tree network [15]. To illustrate our point, we perform simulations on a 3-order 4-layer Cayley tree. As shown in Fig. 3, each link in the Cayley tree has three neighbors. Emanating from link 1, all the links are arranged in shells around vertex 1. In our example there are four such shells. We run ACSMA using the same parameters as in [7] except for that the update interval $T$ is set to 100ms (we assume that a DATA packet duration is 1ms) and $\beta = 5$. Fig. 4 plots $r_i = \log(\rho_i)$, $i = 1, 2$ versus the iteration index, where link 2 is a neighbor of link 1. As can be seen, ACSMA cannot converge. This means that the update interval $T = 100$ ms is not long enough to accurately measure the link throughputs. Then we increase the update interval $T$ by 200 ms each time and repeat the simulation. Finally we obtain that when $T$ is set to 5700ms, ACSMA converges according to our convergence test $\max_j |r_j[n] - r_j^*| / r_j^* < 3\%$ and the number of iteration required is 249. That is, given the update interval $T = 5700$ms, ACSMA needs at least 5700*249 ms $\approx$ 23.66 minutes to converge.

Large update interval $T$ is required to avoid triggering oscillations of $f_i$ and $\log(\rho_i)$ in the Cayley network because the temporal throughputs of links exhibit drastic fluctuations over time. Take link 1 as an example. As plotted in Fig. 5, its normalized temporal throughput alternates between 0 and 1 over time. To exactly measure the throughputs, each link needs to average its measured throughput over several 0-1 cycles, say 8-10 seconds. Thousands of DATA packets are transmitted in each iteration to estimate link throughputs under current network settings. This further slows down the convergence of the algorithm. BP-ACSMA and GBP-ACSMA, however, do not require this real-time measurement and hence will not be affected by this temporal starvation phenomenon.[6]

**A philosophical interpretation of convergence rates**

BP-ACSMA and GBP-ACSMA requires one-hop message passing while ACSMA does not require message passing. One may ponder why BP-ACSMA and GBP-ACSMA can converge faster than ACSMA. A way to look at the problem is as follows. In order for a link $j$ to adjust its access intensity to achieve its fair share of throughput under the utility optimization problem, it somehow has to acquire information about the network topology. To achieve that in a distributed algorithm, the links somehow have to communicate with each other. In BP-ACSMA and GBP-ACSMA, the communication is in the form of "explicit" message passing. The communication in ACSMA, however, is achieved via "implicit messages" in the following sense. In ACSMA, each time a link $j$ transmits a regular data packet, it is actually conveying some information to the neighbor links. In particular, data packets transmitted by link $j$ slow down the clearing of queues in neighbor links, and these links make use of the queue occupancies to adjust their access intensities. Because of the need for smoothing and the fact that these data packets are "indirect" messages, many more data packets than explicit messages are needed in order to convey the same information in ACSMA. This slows down the convergence rate of ACSMA.

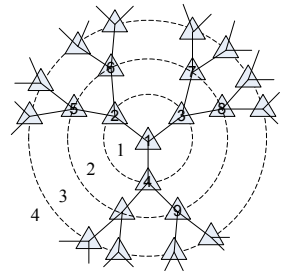

Fig. 3. A three-order Cayley tree network.

The main potential drawback of BP-based algorithms is accuracy, since they only characterize the throughput dependence on the access intensities approximately. More precisely, both BP-ACSMA and GBP-ACSMA are only exact in tree-like topologies (e.g., Cayley tree networks) and may have errors in loopy graphs. The computation error may become unacceptable when the access intensities are extremely large (e.g.,1 e+6) or the network is highly populated. We note, however, that in practice we are unlikely to adopt such large access intensities because of implementation concerns such as finite size of time-slot (see Section III-B of [13], where it was argued that access intensity cannot go beyond 530), higher degree of temporal starvation, etc. For a dense network, we note that GBP-ACSMA can still achieve reasonably accurate results. Section VI details the theory behind GBP and how our specific implementation of GBP for CSMA networks attempts to remove small loops in the message passing construction; small loops are particularly detrimental to accuracy, as we have seen from the example of *N*-vertex circular network discussed in Section III-E.

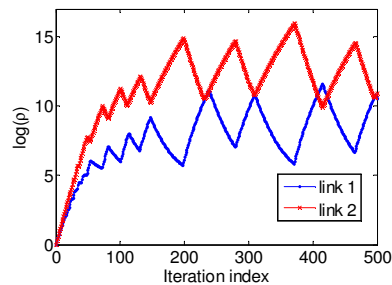

Fig.4. Transmission aggressiveness $r_1$ and $r_2$ of link 1 and link 2 in a 3*4 Cayley tree network for ACSMA of [7] with $T$ = 100ms. Access intensities of other links exhibit similar fluctuations.

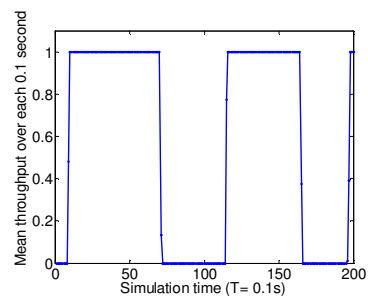

---

[6] We emphasize that we do not claim that BP-ACSMA and GBP-ACSMA can eliminate the temporal starvation phenomenon. Our point is that because BP-ACSMA and GBP-ACSMA obtain the equilibrium throughputs through computation, their convergence will not be slowed down by measurement. All the algorithms studied in this paper focus on controlling the equilibrium throughputs of links. However, given an acceptable equilibrium throughput, the temporal throughput of a link can still alternate between 0 and 1 in cycles of long durations. The reader is referred to [6] on a study on how to characterize temporal starvation and the possible remedies for it.

Fig.5. Normalized throughputs of link 1 in a 3-order 4-layer Cayley tree measured over successive 0.1s intervals when ACSMA is implemented. Throughputs of other links exhibit similar fluctuations.

## VI. GENERALIZED BELIEF PROPAGATION AND ITS APPLICATIONS IN CSMA NETWORKS

In BP, all messages are from one vertex to another vertex. To reduce the error effects of loops, GBP allows messages to be passed from a group of vertices to another group. These groups of vertices are called regions. A region graph is constructed for message passing purposes. The belief of a region corresponds to the joint probability of the states of the vertices within the region. GBP attempts to capture more information than BP because the joint probability of states contains more information on the inter-relationship among the vertices in a region. With the region graph and a new message update rule, GBP can be more accurate than BP.

### A. Region graph

The first step of GBP is to generate a region graph $\mathcal{G}$. In this paper, we use an algorithm similar to the cluster variation method introduced by Kikuchi in 1951 and further developed in the physics literature [16]. The general theory of GBP, however, leaves open the issue of how to define the subsets of vertices to form regions. An important contribution of this paper is to show that a "maximal clique" method of forming regions that are amenable to distributed implementation in CSMA networks yield good results.

A region $R = (V_R, E_R)$ is a subgraph of the original contention graph $G = (V, E)$ in which $V_R \subseteq V$, and $E_R \subseteq E$ are edges between the vertices in $V_R$. Regions are divided into different hierarchical levels. Each region belongs to one of the level. Fig. 6 gives an example demonstrating the construction of a region graph using the cluster variation method.

An important step is the forming of the set of regions at level 0, denoted by $\mathbf{R}_0$. The regions at other levels are constructed based on $\mathbf{R}_0$. That is, the definitions of regions in other levels follow from the definition of $\mathbf{R}_0$. Thus, the definition of $\mathbf{R}_0$ is critical. Every vertex $i \in V$ and every edge $e \in E$ in the original graph must be included into at least one region $R \in \mathbf{R}_0$. We allow for the possibility of a vertex to belong to more than one region in $\mathbf{R}_0$. However, no region $R \in \mathbf{R}_0$ could be a subregion of another region $R' \in \mathbf{R}_0$: that is, $R \not\subset R'$ for any two regions $R, R' \in \mathbf{R}_0$.

In general, there are many ways of forming $\mathbf{R}_0$. Different choices of $\mathbf{R}_0$ correspond to different implementations of GBP. There is a general tradeoff between complexity and accuracy in the choice of $\mathbf{R}_0$. More accuracy can be obtained by GBP if the regions in $\mathbf{R}_0$ are large, but the computation complexity will also be higher.

As discussed in Section III-E, when BP messages are passed around a small loop, computation errors will be incurred. In GBP, we try to include loops in the original graph into a region in $\mathbf{R}_0$ to negate their effects[7]. In our implementation, we generate $\mathbf{R}_0$ by making each maximal clique in $G$ a region in $\mathbf{R}_0$[8]. This ensures that each vertex and each edge in $G$ are included into at least one region. Note in particular that error-inducing small loops in BP consisting of only three vertices are guaranteed to be subsumed into a region in GBP. Although larger loops may not be subsumed into a region, the intuition is that they induce smaller errors anyway. Simulation results in the preceding sections have borne out our method of forming regions in $\mathbf{R}_0$ under various network topologies and parameter settings.

For notational simplicity, in the following we sometimes write $R$ in terms of its vertices only, without listing its edges. In Fig. 6, the maximal cliques are $\{1,2\}$, $\{1,3\}$, $\{3,4\}$, $\{2,4,5\}, \{4,5,6\}$, $\{5,6,8\}$, $\{5,9\}$ and $\{6,7\}$, all of which are included in $\mathbf{R}_0$ on the top row of Fig. 6(b).

After the construction of $\mathbf{R}_0$, we then construct the set of regions at level 1, $\mathbf{R}_1$, from the intersections of the regions in $\mathbf{R}_0$. We discard from $\mathbf{R}_1$, however, any intersection region that is a strict subregion of another intersection region. Specifically, to construct $\mathbf{R}_1$, we first form the set $\mathbf{S}_1 = \{R \mid R = R_i \cap R_j, \forall R_i \in \mathbf{R}_0, R_j \in \mathbf{R}_0, i \neq j\}$. We then discard from $\mathbf{S}_1$ any region $R \in \mathbf{S}_1$ where $R \subset R' \in \mathbf{S}_1$.

In Fig. 6, for example, $\mathbf{R}_1$ consists of $\{1\},\{2\},\{3\},\{4,5\}$ and $\{5,6\}$. Note that although $\{5\}$ is the intersection of $\{2, 4, 5\}$ and $\{5, 6, 8\}$, but $\{5\}$ is not included in $\mathbf{R}_1$ because it is a strict subregion of $\{4, 5\}$ and $\{5, 6\}$.

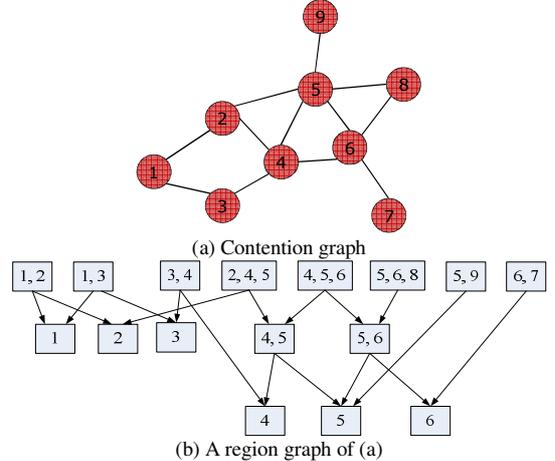

(a) Contention graph

(b) A region graph of (a)

Fig. 6. An example of construction of a region graph.

---

[7] It could be shown that when the resulting region graph does not have a loop, GBP will give exact solutions [16].

[8] It is important to note that the identification of maximal cliques here is not NP-hard if the vertex degree is limited. In practical CSMA wireless networks the degree of a vertex does not grow with the network size, thanks to geographical constraints. Typically, a vertex has at most 5-6 neighbors regardless of the number of vertices in the graph. Let $K$ be the maximum degree of vertices in the contention graph and $N$ be the number of links in the network. For each vertex the complexity of finding maximal cliques containing it is of order $O(2^K)$. Hence, the complexity of finding all the maximal cliques is of order $O(N2^K)$, which increases linearly with $N$. For distributed implementation, the computation-time complexity is of order $O(2^K)$.

Similarly, we construct the set of regions $\mathbf{R}_2$ from the intersections of the regions in $\mathbf{R}_0 \cup \mathbf{R}_1$. In addition to discarding intersection regions that are subregions of other intersection regions in $\mathbf{R}_2$, we also discard intersection regions that have already appeared in $\mathbf{R}_1$.

*General Procedure for Constructing $\mathbf{R}_k$ and Edges to it*

In general, to construct $\mathbf{R}_k$, we first form the set
$$\mathbf{S}_k = \{R \mid R = R_i \cap R_j, \forall R_i \in \mathbf{R}_{k-1}, R_j \in \mathbf{R}_{k-1}, i \neq j\} \cup$$
$$\bigcup_{n=0}^{k-2}\{R \mid R = R_i \cap R_j, \forall R_i \in \mathbf{R}_{k-1}, R_j \in \mathbf{R}_n\}$$

We then discard from $\mathbf{S}_k$ any region $R \in \mathbf{S}_k$ where $R \subset R' \in \mathbf{S}_k$; and any region $R \in \mathbf{S}_k$ where $R \in \mathbf{R}_n$ for some $n \leq k-1$ (i.e., also discard any region in $\mathbf{S}_k$ that already appears at an upper level). We stop forming new regions at the next level when no more new intersection regions can be identified.

For each region $R$, we draw a directed edge from each of its super-regions to it, except for those regions that are super-regions of other super-regions of region $R$. For example, in Fig.6 there is no direct edge from $\{2,4,5\}$ to $\{4\}$, since region $\{2,4,5\}$ is the super-region of region $\{4,5\}$, which is also a super-region of $\{4\}$.

In the resulting region graph $\mathcal{G}$, an edge connects a "parent region" $P$ and a "child region" $R$. If there is a directed path from region $R'$ to region $R$, we say that $R'$ is an ancestor of $R$, and $R$ is a descendant of $R'$. We denote the region graph by $\mathcal{G} = (\mathbf{V}, \mathbf{E})$ where $\mathbf{V}$ is the set of regions and $\mathbf{E}$ is the set of edges. Note that in this paper, to avoid confusion, the bold fonts $\mathbf{V}$ and $\mathbf{E}$ are used to refer to the regions and edges between them, and $V$ and $E$ refers to the vertices and edges between them in the contention graph.

*B. Message and Message-update rules of GBP*

In this paper, we adopt the Parent-to-Child algorithm [17] for message updates. In this algorithm, messages are passed from parent regions to their child regions only. Let $s_R = s_1 s_2 \cdots s_i s_j \cdots s_{|V_R|}$, $i, j \in V_R$ be the state of a region $R$, and $b_R(s_R)$ be the belief of a particular region state $s_R$. In GBP, the "intrinsic" belief of $R$ is given by $\prod_{(i,j) \in E_R} \psi(s_i, s_j) \prod_{i \in V_R} \phi_i(s_i)$. This would be proportional to the probability distribution of the states of the vertices in $R$, if there were no other vertices in the overall network (i.e., if $R$ were the overall network itself). In general, $R$ receives messages from other regions, and these messages capture the correlation of the states of different regions.

Let $\mathbf{D}_R \subseteq \mathcal{G}$ be the subgraph consisting of a region $R$ and all its descendants. In GBP, the update equation of $R$ has to incorporate all "external" messages passed to the regions in $\mathbf{D}_R$, not just those to $R$ only. In Fig. 6, for example, $\mathbf{D}_{\{5,6\}} = \{\{5,6\},\{5\},\{6\}\}$. The following external messages are passed into $\mathbf{D}_{\{5,6\}}$: $m_{\{4,5,6\} \to \{5,6\}}$, $m_{\{5,6,8\} \to \{5,6\}}$, $m_{\{4,5\} \to \{5\}}$, $m_{\{5,9\} \to \{5\}}$, $m_{\{6,7\} \to \{6\}}$.

Let $Parents(R')$ denote the parents of a region $R'$. The belief at $R$ is the product of its intrinsic belief and external messages:
$$b_R(s_R) \propto \prod_{(i,j) \in E_R} \psi(s_i, s_j) \prod_{i \in V_R} \phi_i(s_i) \cdot$$
$$\prod_{R' \in \mathbf{D}_R} \prod_{R'' \in Parents(R') \setminus \mathbf{D}_R} m_{R'' \to R'}(s_{R'}) \quad (19)$$

Note that in the above, the state of $R$ is $s_R$, and the state of $R' \subseteq R$, $s_{R'}$, is induced from $s_R$.

In the *parent-to-child* algorithm, the message-update rules are obtained by requiring consistency of the beliefs between parent and child regions. In Fig. 6(b), let us focus on the region $\{4,5,6\}$ and its child $\{5,6\}$. The belief at region $\{4,5,6\}$ is given by
$$b_{\{4,5,6\}}(s_4 s_5 s_6) \propto \prod_{i,j \in \{4,5,6\}, i \neq j} \psi(s_i, s_j) \prod_{i \in \{4,5,6\}} \phi_i(s_i) m_{\{2,4,5\} \to \{4,5\}}(s_4 s_5) \cdot$$
$$m_{\{5,6,8\} \to \{5,6\}}(s_5 s_6) m_{\{3,4\} \to \{4\}}(s_4) m_{\{5,9\} \to \{5\}}(s_5) m_{\{6,7\} \to \{6\}}(s_6)$$
and the belief at region $\{5,6\}$ is given by
$$b_{\{5,6\}}(s_5 s_6) \propto \psi(s_5, s_6) \phi_5(s_5) \phi_6(s_6) m_{\{4,5,6\} \to \{5,6\}}(s_5 s_6) \cdot$$
$$m_{\{5,6,8\} \to \{5,6\}}(s_5 s_6) m_{\{4,5\} \to \{5\}}(s_5) m_{\{5,9\} \to \{5\}}(s_5) m_{\{6,7\} \to \{6\}}(s_6)$$

Using the marginalization constraint $b_{\{5,6\}}(s_5 s_6) = \sum_{s_4} b_{\{4,5,6\}}(s_4 s_5 s_6)$, we obtain a relation between messages

$$m_{\{4,5,6\} \to \{5,6\}}(s_5 s_6)$$
$$= \frac{\sum_{s_4} \psi(s_4, s_5) \psi(s_4, s_6) \phi_4(s_4) m_{\{2,4,5\} \to \{4,5\}}(s_4 s_5) m_{\{3,4\} \to \{4\}}(s_4)}{m_{\{4,5\} \to \{5\}}(s_5)},$$

which is the *message-update* rule required. Note that on the RHS, with reference to Fig. 6, only those external messages flowing into $\mathbf{D}_{\{4,5,6\}}$ that are not also external messages flowing into $\mathbf{D}_{\{5,6\}}$ are retained in the numerator and only the "internal" messages from $\mathbf{D}_{\{4,5,6\}} \setminus \mathbf{D}_{\{5,6\}}$ to $\mathbf{D}_{\{5,6\}}$ are retained in the denominator. Similar relations can be obtained between each pair of parent and child regions.

In general, the belief of a parent region $P$ can be written as
$$b_P(s_P) \propto \prod_{(i,j) \in E_P} \psi(s_i, s_j) \prod_{i \in V_P} \phi_i(s_i)$$
$$\prod_{P' \in \mathbf{D}_P} \prod_{P'' \in Parents(P') \setminus \mathbf{D}_P} m_{P'' \to P'}(s_{P'}) \quad (20)$$

The marginalization constraint for a child region $R$ with respect to the specific parent $P$ is
$$b_R(s_R) = \sum_{s_{V_{P \setminus R}}} b_P(s_P) \quad (21)$$

Combining (19), (20) and (21), and cancelling common items on the LHS and RHS of (21), the message from a parent $P$ to a child $R$ can be written as

$$m_{P \to R}(s_R) \propto \frac{\sum_{s_{V_{P\setminus R}}} \prod_{(i,j) \in E_P \setminus E_R} \psi(s_i, s_j) \prod_{i \in V_P \setminus V_R} \phi_i(s_i) \prod_{R' \in \mathbf{D}_P \setminus \mathbf{D}_R} \prod_{R'' \in Parents(R') \setminus \mathbf{D}_P} m_{R'' \to R'}(s_{R'})}{\prod_{R' \in \mathbf{D}_R} \prod_{R'' \in Parents(R') \cap \mathbf{D}_P \setminus \mathbf{D}_R} m_{R'' \to R'}(s_{R'})}$$
(22)

Note that the term $\prod_{R' \in \mathbf{D}_P \setminus \mathbf{D}_R} \prod_{R'' \in Parents(R') \setminus \mathbf{D}_P} m_{R'' \to R'}(s_{R'})$ in the numerator consists of the "external" messages into $\mathbf{D}_P$ but not $\mathbf{D}_R$; and the term $\prod_{R' \in \mathbf{D}_R} \prod_{R'' \in Parents(R') \cap \mathbf{D}_P \setminus \mathbf{D}_R} m_{R'' \to R'}(s_{R'})$ in the denominator consists of the "internal" messages from $\mathbf{D}_P \setminus \mathbf{D}_R$ to $\mathbf{D}_R$. Although not necessary mathematically, in each updating we also impose the normalization constraint $\sum_{s_R} m_{P \to R}(s_R) = 1$ to contain the numerical errors.

### C. Distributed GBP

In GBP the messages $m_{P \to R}(s_R)$ are passed from a parent region $P$ to its child region $R$. To implement GBP in a distributed manner, for each message $m_{P \to R}(s_R)$, we need to identify a particular vertex to be responsible for its update and dissemination. We propose to let a vertex that is in both $P$ and $R$, $j \in V_{P \cap R}$, to be such a vertex. Note that $V_{P \cap R}$ could contain more than one vertex. In this case, we elect the vertex with the lowest ID to be the responsible vertex. We will refer to the vertex responsible for a particular message as the *message agent*. As to what to use for ID, we note that each node in the CSMA network usually has a unique ID (e.g., MAC address). Each vertex is a link consisting of a transmitter node and a receiver node. We can simply choose the transmitter node to represent the link, in which case its ID will be the link ID. If we have an infrastructure network, the AP can be chosen to represent the link.

**Features for Correct Operation of Distributed GBP**

The following lists three important features of our distributed GBP that enables its correct operation. These features, which will be proved, mean that each vertex $j$ could deduce its belief $b_j(s_j)$. Details of our distributed GBP will be presented immediately after the description of the features:

**Feature 1:** Each vertex $j$ could collect enough information to construct a local region graph $\mathcal{G}_j$ for the purpose of distributed computation of beliefs and messages. The local region graph $\mathcal{G}_j$ is a subgraph of the complete region graph $\mathcal{G}$. In particular, $\mathcal{G}_j$ is consistent with $\mathcal{G}$ in that each region appearing in $\mathcal{G}_j$ also appears in $\mathcal{G}$, and each edge appearing in $\mathcal{G}_j$ also appears in $\mathcal{G}$.

**Feature 2:** Each vertex $j$ could (i) identify all regions to which it belong from $\mathcal{G}_j$ and randomly select one of them $R$ for its throughput computation; (ii) collect the information needed to compute the region belief $b_R(s_R)$ according to (19). Then, by taking marginal probability, it can compute its throughput: $th_j = b_j(s_j = 1) = \sum_{s_R : s_j = 1} b_R(s_R)$.

**Feature 3:** Each vertex $j$ could (i) identify the messages for which it is the message agent from $\mathcal{G}_j$; and (ii) for each such message $m_{P \to R}(s_R)$, collect the information needed to update $m_{P \to R}(s_R)$ according to (22).

Next, we describe the part of our distributed GBP that enables Feature 1. In our implementation, a vertex $j$ would first construct a local contention graph $G_j$, from which it would construct the local region graph $\mathcal{G}_j$. It does so by listening to the broadcast information from other vertices. We assume that a vertex $j$ can hear the broadcast of all its neighbors $N_j$ in the contention graph $G$.

**Broadcast of All Vertices**

Let $N_j^{(1)} \triangleq N_j \cup \{j\}$. Each vertex $j$ in the network broadcasts three kinds of information in its neighborhood: (i) its link ID $ID_j$; (ii) its access intensity $\rho_j$; (iii) a local contention graph, denoted by $G_j^{(1)}$, consisting of all the vertices in $N_j^{(1)}$ and the edges between them (i.e., all edges $(i,k)$ such that $i,k \in N_j^{(1)}$). Conceptually, this information is embodied in a 3-tuple $(ID_j, \rho_j, G_j^{(1)})$. For ease of exposition, in this paper, we assume $ID_j = j$ and the broadcast information is a 3-tuple $(j, \rho_j, G_j^{(1)})$. The intensity $\rho_j$ is not needed for construction of local contention graphs, and will be used only for the computations of beliefs and messages (to be described in the proofs of Feature 2 and Feature 3). Thus, in the following, we focus on the 2-tuple $(j, G_j^{(1)})$ that can be extracted from the 3-tuple.

**Construction of Local Contention Graph $G_j$**

By assumption, each vertex $j$ could hear the broadcast of all its one-hop neighbors. For each neighbor $i \in N_j$, the broadcast 2-tuple is $(i, G_i^{(1)})$. Vertex $j$ will construct a local contention graph $G_j$ based on $(i, G_i^{(1)})$ from all $i \in N_j$.

Initially $G_i^{(1)} = (i, \varnothing)$ and is not accurate. However, at least all $i \in N_j$ could be identified by vertex $j$ after one round of broadcast by the neighbors. In the next round, each vertex $i \in N_j$, based on what it hears from its neighbors in the last round, can deduce the set of edges $\{(i,k) | k \in N_i^{(1)}\}$. Vertex $i$ will then broadcast $(i, G_i^{(1)})$ with $G_i^{(1)} = (N_i^{(1)}, \{(i,k) | k \in N_i^{(1)}\})$. Specifically, $G_i^{(1)}$ will have the correct vertices, but only edges between $i$ and its neighbors appear; but not those between neighbors. After one more round, however, this will be fixed, and $G_i^{(1)} = (N_i^{(1)}, \{(k,l) | k,l \in N_i^{(1)}, (k,l) \in E\})$ where $E$ are the edges in the complete contention graph $G = (V, E)$. Thus, three rounds of broadcast will make sure the broadcast 2-tuple is correct.

Then, vertex $j$ constructs a local contention graph consisting of the union of its own $G_j^{(1)}$ and the $G_i^{(1)}$ in its neighborhood: $G_j = \bigcup_{i \in N_j^{(1)}} G_i^{(1)}$.

**Property of $G_j$:** $G_j$ contains all vertices within two hops of vertex $j$. From $G_j$, vertex $j$ can identify all maximal cliques to which it belongs (within the overall contention graph $G$), as well as all maximal cliques to which each of its neighbor $i \in N_j$ belongs. That is, all maximal cliques containing at least one vertex in $N_j^{(1)}$ can be identified.

**Construction of Local Region Graph $\mathcal{G}_j$**

Based on $G_j$, vertex $j$ then constructs a local region graph $\mathcal{G}_j$ using the cluster variation method described in Section VI-A, with a small modification, as described in the next paragraph. As in Section VI-A, the first step is to form the set of regions at level 0, denoted by $\mathbf{R}_0^{(j)}$ from the maximal cliques in $G_j$ that contains at least one vertex in $N_j^{(1)}$. After the construction of $\mathbf{R}_0^{(j)}$, we then perform the same procedure as in Section VI-A to construct the regions in lower levels.

The modification is that we will discard all regions that do not contain any vertex in $N_j^{(1)}$ (i.e., $V_R \cap N_j^{(1)} = \varnothing$). The discarded regions will have no bearing on the local computation to be performed. In Fig. 6, for example, let us look at vertex 1. We draw the local contention graph $G_1$ in Fig. 7(a). By forming maximal cliques, $\mathbf{R}_0^{(1)}$ includes regions {1, 2}, {1, 3}, {3, 4} and {2, 4, 5}. At level 1, the intersections of regions in $\mathbf{R}_0^{(1)}$ are generated: {1}, {2}, {3} and {4}. We discard region {4} from level 1 since the sole vertex it contains, vertex 4, is two hops away from vertex 1 and not in $N_j^{(1)}$.

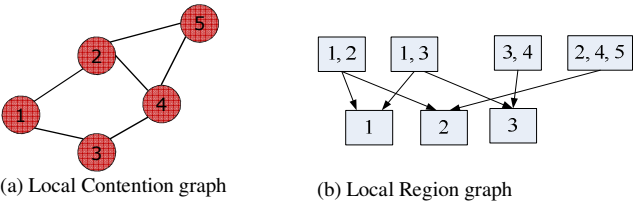

(a) Local Contention graph     (b) Local Region graph

Fig. 7 An example of construction of a local region graph

As in Section IV-A, for each remaining region $R$, we draw a directed edge from each of its super-regions to it, except for those regions that are super-regions of a super-region of region $R$. We denote the local region graph of vertex $j$ by $\mathcal{G}_j = (\mathbf{V}^{(j)}, \mathbf{E}^{(j)})$. Note that $\mathbf{V}^{(j)}$ here are regions and $\mathbf{E}^{(j)}$ are the directed edges between regions.

**Consistency of Local Region Graph**

We next show that the local region graph constructed above is fully consistent with the complete region graph $\mathcal{G}$. We re-state Feature 1 more rigorously here.

**Feature 1:** The local region graph $\mathcal{G}_j$ constructed from $G_j$ is consistent with the complete region graph $\mathcal{G}$ in that each region in $\mathcal{G}_j$ is also a region in $\mathcal{G}$, and each edge in $\mathcal{G}_j$ is also an edge in $\mathcal{G}$. That is, (i) $\forall R \in \mathbf{V}^{(j)}, R \in \mathbf{V}$; (ii) $\forall e \in \mathbf{E}^{(j)}, e \in \mathbf{E}$.

The proof of Feature 1 is given in Appendix C. Based on $\mathcal{G}_j$, we proceed to implement the other procedures of our distributed GBP.

**Selection, Message Computation, and Message Broadcast of Message Agents**

As related earlier, for a message $m_{P \to R}(s_R)$ from a region $P$ to a region $R$, we elect the lowest-ID vertex that is in both $P$ and $R$ to be the message agent responsible for the computation and broadcast of the message. That is, we choose vertex $\arg\min_{i \in V_{P \cap R}}(ID_i)$ to be the message agent for $m_{P \to R}(s_R)$. Feature 2 (proved in Appendix C) implies that vertex $j$ can identify all messages $m_{P \to R}(s_R)$ satisfying $j \in V_{P \cap R}$ from its local region graph $\mathcal{G}_j$. For each such message, vertex $j$ examines the vertices in $V_{P \cap R}$. If it is the vertex with the lowest ID in $V_{P \cap R}$, vertex $j$ will elect itself as the message agent for $m_{P \to R}(s_R)$. It will compute message $m_{P \to R}(s_R)$ according to (22), and then broadcast the message to its neighbors.

We prove Feature 3 in Appendix C that vertex $j$ will be able to collect all the information needed for the computation of $m_{P \to R}(s_R)$. According to (22), other messages may be required for the computation of $m_{P \to R}(s_R)$. We prove that vertex $j$ will be able to hear the broadcast of these messages by their respective message agents (if vertex $j$ is not itself the agent).

**Belief Computation by All Vertices**

Feature 2 states that each vertex $j$ can choose a region $R$ to which it belongs from $\mathcal{G}_j$ and computes the beliefs $b_R(s_R)$ according to (19). It then obtains its throughput by taking marginal probability $th_j = \sum_{s_R: s_j = 1} b_R(s_R)$. Essentially, as with computation of messages, our proof of Feature 2 in Appendix C shows that vertex $j$ will be able to hear the broadcast of the messages required in (19) by their message agents (if vertex $j$ is not itself the agent).

The overall pseudocode of distributed GBP is given below.

**Algorithm 4: Distributed GBP**

1. The following procedure runs on each individual vertex independently. We focus on a particular vertex $j$.
2. Let $N_j^{(1)} = N_j \cup \{j\}$ and $G_j^{(1)}$ be the local contention graph consisting of all the vertices in $N_j^{(1)}$ and the edges between them. Denote the set of the vertices that are within two-hops of vertex $j$ as well as vertex $j$ by $N_j^{(2)}$. Define $G_j = \bigcup_{i \in N_j^{(1)}} G_i^{(1)}$.
3. Let $MS_j$ be the set of messages to which vertex $j$ is the message agent.
4. Vertex $j$ performs the two threads below in parallel.

Thread 1: Periodical Local Information Update

5. Broadcast $(j, \rho_j, G_j^{(1)})$.
6. By listening to the above broadcast of neighbors $N_j$, vertex $j$ derives the local contention graph $G_j$. Using the cluster variation method with the small modification described in Section VI-C, vertex $j$ generates the local region graph $\mathcal{G}_j$.
7. In $\mathcal{G}_j$, for each message $m_{P \to R}(s_R)$ satisfying $j \in V_{P \cap R}$, vertex $j$ examines the vertices in $V_{P \cap R}$. If it is the vertex with the lowest ID in $V_{P \cap R}$, vertex $j$ will elect itself as the message agent for the computation and broadcast of $m_{P \to R}(s_R)$ by adding this $m_{P \to R}(s_R)$ to $MS_j$.
8. Wait for an interval of $T_1$ and repeat the operations of lines 5 and 7, where $T_1$ is an update interval determined by how fast the network contention graph varies (according to the network environment, links leaving and joining the system, etc.).

Thread 2: Message Iteration

9. **procedure** INITIALIZATION
10. $\forall m_{P \to R}(s_R) \in MS_j, \leftarrow \sum_{s_{P \setminus R}} \prod_{i, j \in E_P \setminus E_R} \psi(s_i, s_j) \prod_{i \in V_P \setminus V_R} \phi_i(s_i)$.
11. **end procedure**
12. **procedure** ITERATION
13. broadcast $m_{P \to R}(s_R) \in MS_j$ to all its one-hop neighbors;
14. Wait for time $T_2$ to receive messages from its neighbors, $m_{P \to R}(s_R)$, $\forall P, R \in \mathcal{G}_j$;
15. Based on the received messages and local information maintained by Thread 1, compute each $m_{P \to R}(s_R)$ in $MS_j$ according to (22);
16. Invoke procedure BELIEFCOMPUTATION and repeat procedure ITERATION;
18. **end procedure**
19. **procedure** BELIEFCOMPUTATION
20. Choose any $R$ where $j \in V_R$, compute its belief $b_R(s_R)$ according to (19), and in turn obtain its throughput $th_j$ from marginal probability.
21. **end procedure**

### D. Inverse GBP (IGBP)

Analogous to IBP, we can adapt GBP for the access intensities computation to meet the target throughput distribution.

*1) Message update rules in IGBP*

The first step of IGBP is to construct a region graph using the method introduced in Section VI-A. Second, given $\vec{th}$ (i.e., $b_i(s_i = 1)$ in the BP context), we can obtain $b_R(s_R)$ directly since no more than one link can be active simultaneously in $R$ ($R$ is a clique). If there is any region in $\mathbf{R}_0$ such that the sum of throughputs exceeds 1, we can immediately conclude that the target $\vec{th}$ is not feasible.

Invoking (19) and recall that the links in each region form a clique, we can write

$$\rho_j = \phi_j(s_j = 1) =$$
$$b_R(s_R : s_j = 1) \Big/ \Big( k_R \prod_{R' \in \mathbf{D}_R} \prod_{R'' \in Parents(R') \setminus \mathbf{D}_R} m_{R'' \to R'}(s_{R'}) \Big) \quad (23)$$

where $k_R$ is a normalization factor for $\sum b_R(s_R) = 1$.

Combining (23) with (22), we have the message update rules for IGBP:

i). Based on the messages updated in last iteration and the target throughput, we can solve $\phi_j(s_j = 1)$ for each $j \in V_R$ (i.e., $\rho_j$) from (23);

ii). using $\phi_j(s_j = 1)$ computed in step i), we iterate (22) over each parent-child pair in $\mathcal{G}$.

The iteration stops when $\phi_j(s_j = 1)$ $\forall j$ converges or a maximum number of iterations is reached. Similar to IBP, we cannot guarantee that IGBP converges in general. Recall that $\vec{th} = GBP(G, \vec{\rho})$ is an approximation of the actual link throughputs in the CSMA networks. Similarly, IGBP may output a $\vec{\rho}$ that does not exactly yield the target $\vec{th}$ in the actual network. However, because the computation error has been significantly reduced by forming regions, IGBP has a good chance to approach the target throughputs. Our simulations in Section IV-D validated that the computation error of IGBP is consistently lower than 5% in various networks even if the target throughputs are very close to the upper bound of the capacity region.

*2) Distributed IGBP*

Similar to Distributed GBP, we can also implement IGBP in the distributed manner. We focus on a particular vertex $j$. Besides the local contention graph $G_j$ in distributed GBP, vertex $j$ also has a priori belief at vertex $j$ (i.e., the target throughput $th_j$).

The construction of $G_j$, and in turn $\mathcal{G}_j$, is similar to that of GBP. All vertices, however, need to compute and broadcast access intensities in each iteration, as follows.

**Computation and Broadcast of All Vertices**

Each vertex $j$ in the network computes the access intensity of vertex $j$, $\rho_j$, using (23) and the received messages. After that, it broadcasts the newly computed access intensity $\rho_j$ in its neighborhood.

The pseudocode of IGBP is similar to Algorithm 4. Here, we only show the parts that are different.

---
**Algorithm 5: Distributed IGBP**

In thread 1, remove the broadcast of its access intensity $\rho_j$.

In thread 2, we change from line 10:

10.      $\forall m_{P \to R}(s_R) \in MS_j$, $\leftarrow \sum_{s_{P\setminus R}} \prod_{i,j \in E_P \setminus E_R} \psi(s_i, s_j)$.

14.    Wait for time $T_2$ to receive $\rho_i, \forall i \in N_j$ and messages from its neighbors, $m_{P \to R}(s_R)$, $\forall P, R \in \mathcal{G}_j$;

16.    Invoke procedure ACCESSINTENSITYCOMPUTATION and repeat procedure ITERATION;

17.    Broadcast its access intensity $\rho_j$ to all its one-hop neighbors.

18.    **procedure** ACCESSINTENSITYCOMPUTATION

19.      Compute its access intensity $\rho_j$ according to (23).

20.    **end procedure**

---

*E. GBP-ACSMA*

Analogous to BP-ACSMA, GBP can be adapted for the utility optimization problem as in (11).

*1) Message update rules in GBP-ACSMA*

The first step of GBP-ACSMA is to construct a region graph using the method introduced in Section VI-A.

Second, given the messages in (19), $b_R(s_R)$ can be computed. We can obtain the throughput of a vertex $j$ in region $R$, $th_j$, easily by taking marginal probability from $b_R(s_R)$, exploiting the fact that region $R$ is a clique. Using $th_j$ computed above, we can then solve for $\rho_j$ from (16).

Third, for each message agent $j$, using the newly updated $\rho_j$ in the second step and the received messages, it updates the messages in $MS_j$ according to (22).

*2) Distributed implementation*

Similar to Distributed GBP and IGBP, we can also implement GBP-ACSMA in the distributed manner. We focus on a particular vertex $j$. Besides the local contention graph $G_j$ in distributed GBP, vertex $j$ also has its utility function $U_j(th_j)$.

The construction of $G_j$, and in turn $\mathcal{G}_j$, is similar to that of GBP. All vertices, however, need to compute and broadcast access intensities in each iteration, as follows.

**Computation and Broadcast of All Vertices**

Each vertex $j$ in the network computes the belief of a region to which it belongs according to (19). By taking marginal probability, vertex $j$ gets it throughput $th_j$, using which it solve for $\rho_j$ from (16). In addition, each vertex $j$ broadcasts its newly computed access intensity $\rho_j$ in its neighborhood.

---
**Algorithm 6: GBP-ACSMA**

In thread 1, remove the broadcast of its access intensity $\rho_j$.

In thread 2, we change from line 10:

10.      $\forall m_{P \to R}(s_R) \in MS_j$, $\leftarrow \sum_{s_{P\setminus R}} \prod_{i,j \in E_P \setminus E_R} \psi(s_i, s_j)$.

14.    Wait for time $T_2$ to receive $\rho_i, \forall i \in N_j$ and messages from its neighbors, $m_{P \to R}(s_R)$, $\forall P, R \in \mathcal{G}_j$;

16.    Invoke procedure ACCESSINTENSITYCOMPUTATION and repeat procedure ITERATION;

17.    Broadcast its access intensity $\rho_j$ to all its one-hop neighbors.

18.    **procedure** ACCESSINTENSITYCOMPUTATION

19.      Randomly pick a region $R$ it belongs to, calculate its belief $b_R(s_R)$ using (19). By taking marginal probability, vertex $j$ gets its throughput $th_j$.

20.      Solve for its access intensity $\rho_j$ from (16).

21.    **end procedure**

---

## VII. CONCLUSION

This paper is a first attempt to apply belief propagation to the analysis and design of CSMA wireless networks. In particular, we investigate three applications of belief propagation (BP) and generalized belief propagation (GBP): (1) computation of link throughputs given link access intensities; (2) computation of required link access intensities to meet target link throughputs; and (3) optimization of network utility.

We show how the BP and GBP algorithms for all three applications can be implemented in a distributed manner, making them useful in practical network operation. BP works well in terms of speed, and it yields exact results in tree contention graphs. For loopy contention graphs, GBP can improve accuracy at the cost of longer but still manageable convergence time.

With regard to (1), the problem of computing link throughput given link access intensities in very large CSMA networks is intractable [1]. We show, however, that BP and GBP can obtain accurate approximate results within a short time. This application makes use of the direct correspondence between "beliefs" in the BP framework and "link throughputs" in CSMA networks. In loopy graphs, BP can predict link throughputs with a mean error of less than 10% under various contention-graph and access-intensity settings. GBP can cap the mean error to below 1% for networks of up to 200 links within seconds of computation time.

With regard to (2), we show that the BP framework can be turned around, so that we treat link throughputs as given and compute the link access intensities needed to meet them. This gives rise to the inverse BP and inverse GBP algorithms, in which rather than "belief", it is a network parameter, link access intensity, that gets propagated. Our simulation results show that IBP can output access intensities that give link throughputs that are within 10% of their targets under various contention-graph settings. IGBP can further reduce the difference to below 5%. As for convergence speed, both IBP and IGBP can yield solutions within seconds in real network operation.

Among the three applications, of particular interest are distributed and adaptive algorithms to (3). A solution was first proposed in [7], in which no message passing is needed. The algorithm of [7] is one that is based on "probe and measure". Specifically, before a link adjusts its access intensity in an iteration, a period of "smoothing" time is needed to measure the difference in its input traffic and output traffic of the last iteration. As shown in this paper, the required smoothing time can be quite excessive in networks that exhibit temporal starvation [6], resulting in very slow convergence. BP and GBP adaptive CSMA algorithms, however, do not have this problem because they are computation-based rather than measurement-based. One-hop message passing, however, is required.

Belief propagation has found empirical success in numerous applications (e.g., decoding of LDPC and turbo codes). Typically, the convergence of BP algorithms in these applications is non-trivial to prove (except for tree graphs). Such is the case with belief propagation in CSMA networks as well. For all the scenarios tested, our experiments indicate that both BP and GBP algorithms can converge quickly with accurate computed results. Convergence proofs, however, await future work.

APPENDIX A: INTERPRETATION OF BP IN TREE CONTENTION GRAPH

We now argue that the BP messages propagated in a tree graph can be interpreted as the partition functions of subgraphs. This interpretation reveals why BP can give exact solutions in loop-free graphs.

Consider a vertex $i$ in a tree-like contention graph $G = (V, E)$. Graph $G$ is separated into two subtrees if we remove the edge between $i$ and a neighbor $j$. Let $L(j)$ denote the subtree containing $j$, and let $L^*(j)$ denote the subgraph formed by removing $j$ from $L(j)$. As before, the edges are implied by the existence of vertices.

**Theorem A1:** When applying belief propagation to a tree graph, the message from vertex $j$ to vertex $i$ satisfies

$$m_{ji}(s_i = 0) \propto Z(L(j))$$
$$m_{ji}(s_i = 1) \propto Z(L^*(j))$$
(A1)

where $Z(L(j))$ and $Z(L^*(j))$ are the partition functions of $L(j)$ and $L^*(j)$, respectively.

**Proof:** Let $d_j(v)$ denote the shortest distance (in terms of number of hops) from vertex $v$ to vertex $j$ in the subtree $L(j)$. We prove Theorem A1 by mathematical induction as follows:

1) First we consider the case where $d_j(v) \leq 1 \quad \forall v \in L(j)$. The vertices $v$, if any, are all leaf nodes. If $L^*(j) = \varnothing$, we have

$$m_{ji}(s_i = 0) \propto 1 + \rho = Z(L(j));$$
$$m_{ji}(s_i = 1) \propto 1 = Z(L^*(j)).$$

If $L^*(j) \neq \varnothing$, suppose that $j$ has $n$ one-hop neighbors in $L^*(j)$. We have

$$m_{ji}(s_i = 0) \propto (1+\rho)^n + \rho = Z(L(j))$$
$$m_{ji}(s_i = 1) \propto (1+\rho)^n = Z(L^*(j)).$$

2) Suppose that (A1) holds when $\max_{v \in L(j)} d_j(v) = k$.

When $\max_{v \in L(j)} d_j(v) = k+1$, denote the $n$ neighbors of $j$ by $N_j = \{r_1, r_2, \cdots, r_n\}$. $L(r)$ and $L^*(r)$ are similarly defined for each $r \in N_j$. Note that $\forall v \in L(r)$, we have $v \in L(j)$ and $v$ is connected to $j$ through $r$, so $d_r(v) = d_j(v) - 1$.

Thus we have $\forall r \in N_j$, $\max_{v \in L(r)} d_r(v) \leq k$. Following the precondition above, $\forall r \in N_j$,

$$m_{rj}(s_j = 0) \propto Z(L(r))$$
$$m_{rj}(s_j = 1) \propto Z(L^*(r)).$$

By the message update rule defined in (7), the message from $j$ to $i$ is

$$m_{ji}(s_i = 0) \leftarrow \prod_{r \in N_j} m_{rj}(s_j = 0) + \rho \prod_{r \in N_j} m_{rj}(s_j = 1)$$
$$\propto \prod_{r \in N_j} Z(L(r)) + \rho \prod_{r \in N_j} Z(L^*(r))$$
$$= Z(L(j))$$
$$m_{ji}(s_i = 1) \leftarrow \prod_{r \in N_j} m_{rj}(s_j = 0) \propto \prod_{r \in N_j} Z(L(r)) = Z(L^*(j))$$

Hence, (A1) holds for $\max_{v \in L(j)} d_j(v) = k + 1$.

According to (6), the beliefs at $i$ should be

$$b_i(s_i = 1) \propto \rho \prod_{j \in N_i} m_{ji}(s_i = 1)$$
$$b_i(s_i = 0) \propto \prod_{j \in N_i} m_{ji}(s_i = 0)$$

After normalization, we have $b_i(s_i = 1) =$

$$\rho \prod_{j \in N_i} m_{ji}(s_i = 1) / \left( \rho \prod_{j \in N_i} m_{ji}(s_i = 1) + \prod_{j \in N_i} m_{ji}(s_i = 0) \right)$$
$$= \rho \prod_{j \in N_i} Z(L^*(j)) / \left( \rho \prod_{j \in N_i} Z(L^*(j)) + \prod_{j \in N_i} Z(L(j)) \right) \quad (A2)$$
$$= \rho Z(G - \{i\} \cup N_i) / \left( \rho Z(G - \{i\} \cup N_i) + Z(G - \{i\}) \right)$$

Equation (A2) means that in a tree graph, BP correctly computes $p_i(s_i = 1) = th_i$ of ICN in the form of $Z_i / Z$ expressed in Section II.

APPENDIX B: PROOF OF THE THEOREM 1

To prove Theorem 1, we first present a simplified belief propagation (SBP) that is equivalent to the original BP in CSMA networks. Using SBP, we show the convergence of IBP.

A. *Simplified belief propagation (SBP)*

In the body of the paper, we use two variables to express the messages and beliefs. As mentioned there, the beliefs need to be normalized so that $\sum_{s_i \in \{0,1\}} b_i(s_i) = 1$ and $\sum_{s_i \in \{0,1\}} m_{ji}(s_i) = 1$. That is, in message passing, only the "ratios" are useful. Noting that in a finite CSMA network, the belief of a link state cannot be either 1 or 0 according to the ICN model. We define

$$n_{ij} = \frac{m_{ij}(s_j = 1)}{m_{ij}(s_j = 0)} \quad (B1)$$

and

$$c_i = \frac{b_i(s_i = 1)}{b_i(s_i = 0)} \quad (B2)$$

Instead of dealing with $b_i(s_i = 0)$, $b_i(s_i = 1)$, $m_{ij}(s_j = 0)$, and $m_{ij}(s_j = 1)$, we can deal with $c_i$ and $n_{ij}$. Accordingly, the update rules of BP are revised as follows:

$$\left. \begin{array}{l} b_i(s_i = 0) = k_i \prod_{j \in N_i} m_{ij}(s_i = 0) \\ b_i(s_i = 1) = k_i \rho_i \prod_{j \in N_i} m_{ij}(s_i = 1) \end{array} \right\} \Rightarrow c_i = \rho_i \prod_{j \in N_i} n_{ji} \quad (B3)$$

From (7), we have

$$\left. \begin{array}{l} m_{ji}(s_i = 0) = \frac{b_j(s_j = 0)}{k_j m_{ij}(s_j = 0)} + \frac{b_j(s_j = 1)}{k_j m_{ij}(s_j = 1)} \\ m_{ji}(s_i = 1) = \frac{b_j(s_j = 0)}{k_j m_{ij}(s_j = 0)} \end{array} \right\} \Rightarrow n_{ji} = \frac{n_{ij}}{n_{ij} + c_j}$$
(B4)

Equations (B3) and (B4) form a simpler update rule to perform belief propagation. The number of equations is reduced by half.

B. *Proof the Theorem 1*

In IBP, the belief of each link $b_j(s_j)$ is given from the target throughput. That is, $c_j$ defined in (B2) is pre-fixed in IBP and there is no need to update it. In SBP, the message update rule is (B4). It iterates (B4) over all vertices $j$, and the desired link access intensity $\rho_j$ is obtained from

$$\rho_j = \frac{c_j}{\prod_{i \in N_j} n_{ij}} \quad (B5)$$

Recall that only "ratios" are useful in belief propagation, the simplified IBP defined by (B4) and (B5) is similar to the IBP defined by (8) and (9) in nature. We next investigate the convergence of the simplified IBP.

If the target throughput of IBP is feasible in the sense that $\vec{th} = BP(G, \vec{\rho})$ for some $\vec{\rho}$, then the desired output of IBP should be $\vec{\rho}$. In message update, we denote the corresponding messages by $n_{ji}^*$. That is, $n_{ji}^*$ is the converged message if the algorithm converges correctly.

Consider any pair of vertices which perform the IBP procedure. Let $n_{ji}^{(k)}$ be the message from vertex $j$ to vertex $i$ in the $k^{th}$ iteration. In the $(k+1)^{th}$ iteration, the messages computed are

$$n_{ij}^{(k+1)} = \frac{n_{ji}^{(k)}}{n_{ji}^{(k)} + c_i}$$

and

$$n_{ji}^{(k+1)} = \frac{n_{ij}^{(k+1)}}{n_{ij}^{(k+1)} + c_j} = \frac{n_{ji}^{(k)}}{n_{ji}^{(k)} + c_j n_{ji}^{(k)} + c_i c_j} \quad (B6)$$

We look at the distance between $n_{ji}^{(k)}$ and $n_{ji}^*$. Write $\Delta n_{ji}^{(k)} = \left| n_{ji}^{(k)} - n_{ji}^* \right|$. To show that the message computed in IBP converges, it is sufficient to show that

$$\Delta n_{ji}^{(k+1)} \leq \varepsilon \Delta n_{ji}^{(k)} \text{ for some } \varepsilon < 1 \quad (B7)$$

That is, the messages iterated in IBP is a contraction mapping and guaranteed to converge to the fixed point $n_{ji}^*$.

The following shows (B7):

Because $n_{ji}^*$ is the desired fixed point, we have

$$n_{ji}^* = \frac{n_{ji}^*}{n_{ji}^* + c_j n_{ji}^* + c_i c_j} \quad (B8)$$

$$\Delta n_{ji}^{(k+1)} = \left| n_{ji}^{(k+1)} - n_{ji}^* \right| = \left| \frac{n_{ji}^{(k)}}{n_{ji}^{(k)} + c_j n_{ji}^{(k)} + c_i c_j} - \frac{n_{ji}^*}{n_{ji}^* + c_j n_{ji}^* + c_i c_j} \right|$$

$$= \frac{c_i c_j \left| n_{ji}^{(k)} - n_{ji}^* \right|}{\left( n_{ji}^{(k)} + c_j n_{ji}^{(k)} + c_i c_j \right)\left( n_{ji}^* + c_j n_{ji}^* + c_i c_j \right)} \quad (B9)$$

$$= \frac{c_i c_j \Delta n_{ji}^{(k)}}{\left( n_{ji}^{(k)} + c_j n_{ji}^{(k)} + c_i c_j \right)\left( n_{ji}^* + c_j n_{ji}^* + c_i c_j \right)}$$

From (B8), we know that $n_{ji}^* + c_j n_{ji}^* + c_i c_j = 1$. Furthermore, $n_{ji}^{(k)} + c_j n_{ji}^{(k)} > 0$. Thus,

$$\frac{c_i c_j}{\left( n_{ji}^{(k)} + c_j n_{ji}^{(k)} + c_i c_j \right)\left( n_{ji}^* + c_j n_{ji}^* + c_i c_j \right)} < 1 \quad (B10)$$

APPENDIX C: DISTRIBUTED IMPLEMENTATION OF GBP

This section proves the three features for correct operation of Distributed GBP. We first review and summarize some properties of the region graph $\mathcal{G}$ constructed using the method in Section VI-A, as well as some properties of the local contention graph $G_j$ and local region graph $\mathcal{G}_j$ constructed using the method in Section VI-C. These properties will be used in our proofs.

**Property 1:** All maximal cliques in $G$ that contain at least one vertex in $N_j^{(1)}$ can be identified from $G_j$. Each region in $\mathbf{R}_0^{(j)}$ of $\mathcal{G}_j$ is one of these maximal cliques. In addition, $\mathbf{R}_0$ of $\mathcal{G}$ contains all these regions.

**Property 2:** All regions in $\mathcal{G}$ and $\mathcal{G}_j$ are cliques.

**Property 3:** For two regions $R$ and $R'$ in $\mathcal{G}$ ($\mathcal{G}_j$), if $R \cap R' \neq \emptyset$, then $R \cap R'$ is also a region in $\mathcal{G}$ ($\mathcal{G}_j$ if $R \cap R'$ contains at least one vertex in $N_j^{(1)}$).

**Property 4:** Consider two regions $R$ and $R'$ in $\mathcal{G}$ ($\mathcal{G}_j$) such that $R' \subset R$. There is a direct edge from $R$ to $R'$ if and only if there does not exist another region $R''$ in $\mathcal{G}$ ($\mathcal{G}_j$) such that $R' \subset R'' \subset R$.

**Property 5:** Any region in $\mathcal{G}$ that contains at least one vertex in $N_j^{(1)}$ must also be a region in $\mathcal{G}_j$.

Property 1 is a property of $G_j$, which has been elaborated in the construction of $G_j$ in Section VI-C. Properties 2 and 4 are directed consequences of our region graph construction method described in Section VI-A and Section VI-C.

We give an explanation to Property 3 with respect to $\mathcal{G}$ as follows (similar explanation applies to $\mathcal{G}_j$ because it uses the same construction method). In the construction of $\mathcal{G}$, all regions except those in $\mathbf{R}_0$ are generated by the intersections of regions at the upper levels. Although we discard some regions during the construction (see Section VI-A, in which it was mentioned that "We then discard from $\mathbf{S}_k$ any region $R \in \mathbf{S}_k$ where $R \subset R' \in \mathbf{S}_k$; and any region $R \in \mathbf{S}_k$ where $R \in \mathbf{R}_n$ for some $n \leq k-1$…"), we note these discarded regions either already exist at a higher level, or will be added back at a lower level.

To see Property 5, consider a region $R$ in the complete region graph $\mathcal{G}$ containing at least one vertex in $N_j^{(1)}$. If $R$ is a maximal clique (i.e., $R \in \mathbf{R}_0$), then by Property 1, $R \in \mathbf{R}_0^{(j)}$. If $R$ is not a maximal clique, then $R$ has at least two ancestors in $\mathbf{R}_0$, $R', R'' \in \mathbf{R}_0$, containing a vertex in $N_j^{(1)}$. By Property 1, $R', R'' \in \mathbf{R}_0^{(j)}$. Thus, by Property 3, $R \in \mathbf{V}^{(j)}$. In summary, $R \in \mathcal{G}_j$.

**Proof of Feature 1:** We first prove (i). First, consider the regions at level 0. By Property 1, all $R \in \mathbf{R}_0^{(j)}$ must also be in $\mathbf{R}_0$. A region $R$ at a lower level of $\mathcal{G}_j$ ($\mathcal{G}$) is generated from the intersection of the regions at the upper layers. In particular, lower-level regions are induced by the regions at level 0. By Property 3, each region $R$ in $\mathbf{V}^{(j)}$ at levels below level 0 must also be a region in $\mathbf{V}$. That is, $\forall R \in \mathbf{V}^{(j)}, R \in \mathbf{V}$. The local region graph $\mathcal{G}_j$ does not include any extraneous regions not in $\mathcal{G}$.

We prove (ii) by contradiction. Suppose that there is a pair of parent-child regions $R, R'$ in $\mathcal{G}_j$ with an edge between them in $\mathcal{G}_j$ but no edge between them in $\mathcal{G}$. Without loss of generality, let $R$ be the parent, (i.e, $R' \subset R$). Invoking Property 4, there must exist a region $R'' \in \mathbf{V}$ and $R'' \notin \mathbf{V}^{(j)}$, such that $R' \subset R'' \subset R$. Note that by our construction method for $\mathcal{G}_j$, $R'$ must contain at least one vertex in $N_j^{(1)}$. Together with $R' \subset R''$, this means $R''$, which is not in $\mathcal{G}_j$, must have at least one vertex in $N_j^{(1)}$. This contradicts Property 5.

Before we proceed to prove Features 2 and 3, we put down an extra property of our region graph.

**Property 6:** Consider two regions $R$ and $R'$ in $\mathcal{G}$ between which there is an edge. If $R$ and $R'$ are also regions in $\mathcal{G}_j$, there must be an edge between them as well in $\mathcal{G}_j$.

**Proof of Property 6:** Without loss of generality, we assume $R$ is the parent of $R'$. Invoking Property 4, there does not exist another region $R''$ in $\mathcal{G}$ such that $R' \subset R'' \subset R$. Suppose that $R$ and $R'$ also exist in $\mathcal{G}_j$ and there is no edge between them in $\mathcal{G}_j$. Invoking Property 4, there must be another region $R''$ in $\mathcal{G}_j$ such that $R' \subset R'' \subset R$. However, according to Feature 1, $\mathcal{G}_j$ does not contain any extraneous regions not in $\mathcal{G}$. Thus, the existence of the extraneous region $R''$ cannot be true. □

Feature 1' below combines Properties 5 ad 6 to facilitate articulation of the proofs of Features 2 and 3 later:

**Feature 1':** Any region $R$ in $\mathcal{G}$ that contains at least one vertex in $N_j^{(1)}$ must also be a region $R$ in $\mathcal{G}_j$. Consider two regions $R$ and $R'$, both having at least one vertex in $N_j^{(1)}$. If there is an edge between $R$ and $R'$ in $\mathcal{G}$, there is also an edge between $R$ and $R'$ in $\mathcal{G}_j$.

**Comment:** Recall that Feature 1 means there are no extraneous regions or extraneous edges between regions in $\mathcal{G}_j$. Feature 1' is sort of a converse to Feature 1. As will be seen, it means that the portion of the region graph structure in $\mathcal{G}$ needed for the computation of local beliefs and messages by vertex $j$ is exactly duplicated in $\mathcal{G}_j$.

We next prove Features 2 and 3.

**Proof of Feature 2:** According to Feature 1, vertex $j$ has a local region graph $\mathcal{G}_j$ with no extraneous vertices or edges absent in $\mathcal{G}$. According to Feature 1', all regions in $\mathcal{G}$ that contain vertex $j$ must also be in $\mathcal{G}_j$. Thus, vertex $j$ could identify all the regions in $\mathcal{G}$ to which it belongs. For belief computation, vertex $j$ could choose a small region $R$ among such regions (for computation simplicity). For vertex $j$ to compute $b_R(s_R)$ as per (19), it needs the following information: (a) $\psi(s_i, s_k)$, $\forall(i,k) \in E_R$ (note: we change the index $j$ in (19) to $k$ here to avoid confusion with vertex $j$ here) and $\phi_i(s_i) \forall i \in V_R$, and (b) the messages from external regions into $\mathbf{D}_R$.

(a) is trivial because all $i \ne j$ that are in $R$ are one-hop neighbors of $j$. Thus, the broadcast of their access intensities $\phi_i(1) = \rho_i$, as described in Section VI-C, can be heard by vertex $j$; and $\phi_i(0) = 1$ by definition. As to $\psi(s_i, s_k)$, it is already available by definition: $\psi(s_i, s_k) = 0$ if $s_i = s_k = 1$; $\psi(s_i, s_k) = 1$ otherwise.

(b) needs to be further separated into two steps. Vertex $j$ needs to be able to (i) identify the external messages for $\mathbf{D}_R$, and (ii) hear them when their agents broadcast them. An external message passed into a region $R' \in \mathbf{D}_R$ is of the form $m_{P' \to R'}(s_{R'})$, where $P'$ is a parent region of $R'$ not within $\mathbf{D}_R$.

For (i), we note that $R' \subseteq R$ because $R'$ is either $R$ or a descendant of $R$. Thus, all vertices in $R'$ are one-hop neighbors of vertex $j$. Combining with $R' \subset P'$, we deduce that $P'$ must have at least one vertex that is in $N_j^{(1)}$. Invoking Feature 1', we have that both $P'$ and $R'$ are regions in $\mathcal{G}_j$, and there is an edge from $P'$ to $R'$ in $\mathcal{G}_j$. Since there are no extraneous regions and edges in $\mathcal{G}_j$ either (Feature 1), vertex $j$ will be able to correctly deduce the portion of the graph structure of $\mathcal{G}$ relevant to the computation of $b_R(s_R)$ (i.e., this portion is exactly duplicated in $\mathcal{G}_j$).

For (ii), according to our distributed implementation, a vertex $i \in R'$ is chosen as the agent for the computation and broadcast of $m_{P' \to R'}(s_{R'})$. Since all vertices in $R'$ are one-hop neighbor of vertex $j$, vertex $j$ can hear $m_{P' \to R'}(s_{R'})$.

**Proof of Feature 3:** As mentioned in the first paragraph of the proof of Feature 2, vertex $j$ could identify all regions in $\mathcal{G}$ to which it belongs. Suppose that vertex $j$ belongs to two regions $P$ and $R$. According to Features 1 and 1', the presence or absence of an edge between $P$ and $R$ is exactly duplicated in $\mathcal{G}_j$. If there is an edge, vertex $j$ will be able to decide whether it should be the message agent for the edge (according to our implementation, vertex $j$ will elect itself as the message agent if it is the vertex with the lowest ID in $V_{P \cap R}$). If there is no such edge $\mathcal{G}_j$, then there is no such edge in $\mathcal{G}$, and vertex $j$ will not miss out any message for which it is responsible.

Consider a message $m_{P \to R}(s_R)$ to which vertex $j$ is the agent. For vertex $j$ to compute each $m_{P \to R}(s_R)$ as per (22), it needs the following information: (a) $\psi(s_i, s_k)$, $\forall(i,k) \in E_P \setminus E_R$ and $\phi_i(s_i)$, $\forall i \in V_P \setminus V_R$, (b) messages $m_{R'' \to R'}(s_{R'})$, $R' \in \mathbf{D}_P \setminus \mathbf{D}_R$ and $R'' \in Parents(R') \setminus \mathbf{D}_P$, and (c) messages $m_{R'' \to R'}(s_{R'})$, $R' \in \mathbf{D}_R$ and $R'' \in Parents(R') \cap \mathbf{D}_P \setminus \mathbf{D}_R$.

(a) is trivial. We note that $j \in P$ and $j \in R$. And the rest of the argument is the same as the proof for (a) related to Feature 2.

The arguments for (b) and (c) are also similar to the argument for (b) in the proof of Feature 2. Essentially, the portion of the graph structure in $\mathcal{G}$ relevant to the messages in (b) and (c) are exactly duplicated in $\mathcal{G}_j$, so that vertex $j$ can identify these messages properly. Furthermore, the agents for these messages must be either vertex $j$ itself or one-hop neighbors of vertex $j$.

APPENDIX D: FIXED POINT OF BP IN THE RING GRAPH

We prove that BP converges to the fixed point $b_i(s_i = 0) = (1 + \sqrt{1+4\rho})/2\sqrt{1+4\rho}$ for each vertex in any $N$-vertex ring graph regardless of $N$.

**Proof:**

By symmetry, in each iteration, the messages being passed from $i$ to $j$ are the same for all pairs of neighbors $i, j$. Let the vector $\begin{pmatrix} m^{(n)}(0) \\ m^{(n)}(1) \end{pmatrix}$ denote the message being passed in iteration $n$. We omit the subscripts $i, j$ in our notation because all messages are the same. Applying (7) on this ring contention graph, we get the following dynamic equation for messages:

$$\begin{pmatrix} m^{(n)}(0) \\ m^{(n)}(1) \end{pmatrix} \propto \begin{pmatrix} \phi(0) & \phi(1) \\ \phi(0) & 0 \end{pmatrix} \begin{pmatrix} m^{(n-1)}(0) \\ m^{(n-1)}(1) \end{pmatrix} = \begin{pmatrix} 1 & \rho \\ 1 & 0 \end{pmatrix} \begin{pmatrix} m^{(n-1)}(0) \\ m^{(n-1)}(1) \end{pmatrix} \quad (C1)$$

From (C1), we can get

$$\begin{aligned} m^{(n)}(0) &= m^{(n-1)}(0) + \rho m^{(n-2)}(0) \\ m^{(n)}(1) &= m^{(n-1)}(0) \end{aligned} \quad (C2)$$

The solution to the difference equation (the first equation in (C2)) is

$$m^{(n)}(0) = Cz_1^n + Dz_2^n \quad (C3)$$

where $z_1, z_2 = \dfrac{1 \pm \sqrt{1+4\rho}}{2}$, and $C$ and $D$ are constants to match the boundary condition.

For example, if $\begin{pmatrix} m^{(0)}(0) \\ m^{(0)}(1) \end{pmatrix}$ has been initialized to $\begin{pmatrix} 1 \\ 1 \end{pmatrix}$. Then

$$\begin{aligned} C + D &= 1 \\ Cz_1 + Dz_2 &= 1 + \rho \end{aligned}$$

which gives

$$\begin{aligned} C &= \frac{1+\rho-z_2}{z_1-z_2} = \frac{\frac{1}{2}+\rho+\frac{1}{2}\sqrt{1+4\rho}}{\sqrt{1+4\rho}} = \frac{\frac{1}{2}+\rho}{\sqrt{1+4\rho}} + \frac{1}{2} \\ D &= \frac{-\frac{1}{2}-\rho+\frac{1}{2}\sqrt{1+4\rho}}{\sqrt{1+4\rho}} = -\frac{\frac{1}{2}+\rho}{\sqrt{1+4\rho}} + \frac{1}{2} \end{aligned} \quad (C4)$$

The belief is given by

$$\begin{aligned} b^{(n)}(0) &= \frac{\phi(0)m^{(n)}(0)^2}{\phi(0)m^{(n)}(0)^2 + \phi(1)m^{(n)}(1)^2} \\ &\to \frac{z_1^{2n}}{z_1^{2n} + \rho z_1^{2(n-1)}} = \frac{1}{1+\rho\left(\dfrac{1+\sqrt{1+4\rho}}{2}\right)^{-2}} \\ &= \frac{1+2\rho+\sqrt{1+4\rho}}{1+4\rho+\sqrt{1+4\rho}} = \frac{\left(1+\sqrt{1+4\rho}\right)^2}{2\sqrt{1+4\rho}\left[1+\sqrt{1+4\rho}\right]} \\ &= \frac{\left(1+\sqrt{1+4\rho}\right)}{2\sqrt{1+4\rho}} \end{aligned} \quad (C5)$$